\documentclass[showpacs,prb,amsmath,amssymb]{revtex4}
\usepackage{graphicx}
\usepackage{color}
\usepackage{bm}
\usepackage{psfrag}
\long\def\comment#1{}

\begin{document}

\title{General topological features and instanton vacuum in quantum Hall
and spin liquids}
\author{A.M.M.~Pruisken}
\affiliation{Institute for Theoretical Physics, University
of Amsterdam, Valckenierstraat 65, 1018 XE Amsterdam, The
Netherlands.}
\author{R.~Shankar}
\author{Naveen Surendran\footnote{Present Address: {\em Institute
for Theoretical Physics, University of Amsterdam,
Valckenierstraat 65, 1018 XE Amsterdam, \mbox{The Netherlands.}} }}
\affiliation{The Institute of Mathematical
Sciences, CIT Campus, Chennai 600 113, India.}

\begin{abstract}
We introduce the concept of {\em super universality} in quantum
Hall liquids and spin liquids. This concept has emerged from
previous studies of the quantum Hall effect and states that all
the fundamental features of the quantum Hall effect are generically
displayed as {\em general topological features} of the
$\theta$ parameter in nonlinear $\sigma$ models in two dimensions.

To establish super universality in spin liquids we revisit the
mapping by Haldane who argued that the anti ferromagnetic
Heisenberg spin $s$ chain in $1+1$ space-time dimensions is
effectively described by the $O(3)$ nonlinear $\sigma$ model with
a $\theta$ term. By combining the path integral representation for
the dimerized spin $s=1/2$ chain with renormalization group
decimation techniques we generalize the Haldane approach to
include a more complicated theory, the {\em fermionic rotor
chain}, involving four different renormalization group parameters.
We show how the renormalization group calculation technique can be
used to lay the bridge between the {\em fermionic rotor chain} and
the $O(3)$ nonlinear $\sigma$ model with the $\theta$ term.

As an integral and fundamental aspect of the mapping we establish
the topological significance of the {\em dangling spin} at the
edge of the chain. The edge spin in spin liquids is in all respects
identical to the {\em massless chiral edge excitations} in quantum Hall
liquids. We consider various different geometries of the spin chain such
as open and closed chains, chains with an {\em even} and {\em odd}
number of sides. We show that for each of the different geometries
the $\theta$ term has a distinctly different physical meaning. We
compare each case with a topologically equivalent quantum Hall
liquid.

\end{abstract}

\maketitle

\section{\label{Intro} Introduction}
The topological concept of an instanton vacuum in scale invariant
theories is one of the outstanding strong coupling problems in
physics that to date has generally not been understood
~\cite{Raja}. Perhaps the most interesting and profound
application of this concept is found in the theory of the {\em
Quantum Hall Effect} (QHE)~\cite{Prange}. The otherwise somewhat
obscure {\em instanton angle} $\theta$ has a clear physical
significance in this case. It represents a physical observable,
namely the {\em Hall conductance}.

The advances\cite{PruiskenBaranovVoropaev,PruiskenBurmistrov}
made over many years have resulted in a clear physical
understanding of the $\theta$ dependence in the Grassmannian
$U(M+N)/U(M)\times U(N)$ non-linear $\sigma$ model (NLSM) in terms
of quantum Hall physics. In dramatic contrast to the
standard lore which states that in this problem the {\em replica
limit} (i.e. $M,N \rightarrow 0$) is all-important, one can now
say that the fundamental features of the QHE are all displayed as
{\em super universal topological features} of the {\em instanton
vacuum}, independent of the details such as the number of field
components ({\em replica's}) $M$ and $N$ in the theory. The list
of super universal topological features includes the existence of
{\em gapless excitations} at $\theta=\pi$, the appearance of {\em
massless chiral edge excitations} as well as the existence of {\em
robust topological quantum numbers} that explain the {\em
observability} and {\em precision} of the QHE. These fundamental
strong coupling aspects of the instanton vacuum should not
be confused with the concept of {\em ordinary universality}.
Ordinary universality applies solely to the details of the
critical behaviour at $\theta=\pi$ which may in principle vary as
one varies the number of field components in the theory. Following
the standard phenomenology of critical phenomena
one may associate different universality classes with
different values of $M$ and $N$.
\begin{figure}\label{RGflowMN}
\includegraphics[height=320pt]{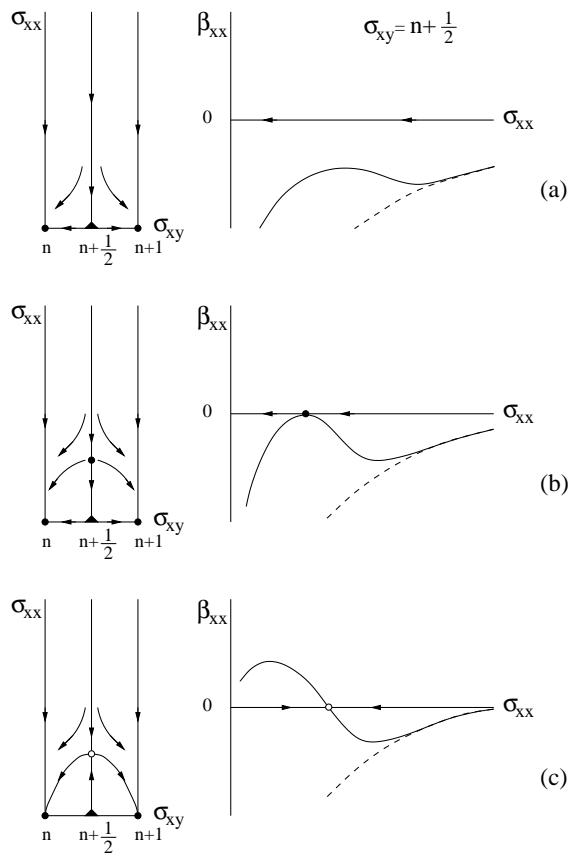}
\caption{ The renormalization group flow diagram for different
values of $M$ and $N$ in the $\sigma_{xx}$, $\sigma_{xy}$
conductance plane. Here, $\sigma_{xx} =1/2g$ and $\sigma_{xy} =
\theta/2\pi$ where $g$ denotes the coupling constant and $\theta$
the instanton angle. (a) The unstable fixed points along the lines
$\sigma_{xy} = n+1/2$ may have a critical value $\sigma_{xx}^* =0$
as found in the large $N$ expansion of the $CP^{N-1}$ model or,
equivalently, $SU(N)/U(N-1)$. (b) The value $\sigma_{xx}^*$ is
finite but with a {\em marginally irrelevant} direction
($\sigma_{xx} > \sigma_{xx}^*$) and a {\em marginally relevant}
($\sigma_{xx} > \sigma_{xx}^*$) direction respectively. This
behaviour is expected for $M=N=1$ or $SU(2)/U(1) \backsimeq O(3)$.
(c) The critical point has a finite $\sigma_{xx}^*$ and is stable
along the entire $\sigma_{xy} = n+1/2$ axis. This behaviour is
typical for the theory with a small number of field components $1
\gtrsim M,N \geqq 0$.}
\end{figure}
The various different aspects of the Grassmannian
$U(M+N)/U(M)\times U(N)$ theory with varying values of $M$ and $N$
are encapsulated in the renormalization group flow diagrams of
Fig.~\ref{RGflowMN}. The {\em super universal} features are
represented by the infrared stable fixed points at $\theta=2\pi
k$, as well as the the unstable fixed points at $\theta =
\pi(2k+1)$ which generally describe {\em gap-less} excitations or
a {\em divergent} correlation length in the bulk of the system.

It is important to emphasize that the discovery of massless edge
excitations has a major impact on our general understanding of the
instanton vacuum at strong coupling ~\cite{Unify3,Unify4}.
It has resulted in a fundamental
revision~\cite{PruiskenBaranovVoropaev} of commonly accepted but
conflicting ideas and expectations that were all based on extended
previous work on an ``exactly" solvable limit of the theory, the
large $N$ expansion of the $CP^{N-1}$ model~\cite{LargeN2}. These
historical analyzes have set the stage for the $\theta$ dependence
in an overwhelming majority of Grassmannian $U(M+N)/U(M) \times
U(N)$ NLSM's, presumably all integers $M$ and $N$ larger than
unity, for which the physics at large distances is likely to be
the same. Assuming the historical results and claims to be true,
however, then none of the aforementioned super universal
features of the instanton vacuum would actually exist. This
dramatic conflict between the historical claims and the physics of
the QHE has spanned the subject of an instanton parameter $\theta$
for a very long time. It is just one of the reasons why it has
proven to be so difficult in this field to pursue the right
physical ideas as well as the appropriate mathematical questions.

It so turned out that much of the historical large $N$ analysis is
fundamentally incorrect~\cite{PruiskenBaranovVoropaev}. For one
thing, the topological subtleties of the ``edge" in this problem
have been entirely overlooked and, along with that, the most
relevant part of the excitation spectrum, namely the {\em
gap-less} bulk excitations that generally exist at $\theta=\pi$.
Much of the dilemma is due to the fact that although the
topological phase transition at $\theta=\pi$ may formally be a
{\em first order} one, the system can nevertheless exhibit a
diverging correlation length and, hence, display all the
characteristics of a {\em continuous, second order transition}. In
any case, in dramatic contrast to what the historical analyzes
were saying or trying to say, the large $N$ expansion, as it now
stands, is one of the very rare and interesting examples where the
super universal strong coupling features of the instanton vacuum
concept can be fully explored and demonstrated in great detail.

In this paper we investigate the {\em super universality} concept
in a somewhat different physical context, the one dimensional
anti-ferromagnetic Heisenberg spin chain (HSC). In 1983 Haldane
\cite{haldane} argued that the HSC, in the limit of large spin
$s$, is effectively described by the $SU(2)/U(1) \backsimeq O(3)$
NLSM with a $\theta$ term. On the basis of this mapping (in brief
{\em Haldane mapping}) Haldane predicted that the HSC with $s=1$
has a mass gap whereas for $s=1/2$ the chain is known to be
gapless. Subsequently, Affleck \cite{affleck} extended the
argument to include the {\em dimerized Heisenberg spin chain}
(DSC). For the DSC the parameter $\theta$ becomes, like in the
theory of the QHE, a continuously varying parameter. The relevant
variable in the case is the so-called {\em dimerization parameter}
$\kappa$.

As a general remark we can say that Haldane conjecture for spin
chains naturally emerges from the renormalization group flow
diagram as it was obtained from the studies of the QHE
(Fig.~\ref{RGflowMN}b). In so far as one can trust the Haldane
mapping for more complicated situations such as the DSC for
arbitrary $s$, dimerized $SU(N)$ spin chains etc., one proceed
along similar lines and obtain the global phase diagram of the
spin system from the flow diagram of the corresponding NLSM.

Unfortunately, the Haldane mapping has in general not been
understood well enough to facilitate an unambiguous, one-to-one
mapping between the spin chain and continuum field theory, i.e.
the NLSM in the presence of the $\theta$ term. While the theory
near the $\theta=\pi ~(\kappa=1)$ has been extensively studied in
the context of both the DSC \cite{rrps} and the $O(3)$ NLSM
\cite{mussardo} nobody has as of yet recognized that both models
display all the fundamental strong coupling features of the QHE.
The complications in the Haldane mapping are clearly reflected by
the fact that even to date some unresolved issues have remained
such as the difference between spin chains with an {\em even} and
an {\em odd} number of spins. A related problem is the peculiar
role played by the {\em dangling spin} at the edge of the chain.
In all these cases it is not quite understood how the Haldane
mapping should be carried out, or what the $\theta$ term in the
$O(3)$ NLSM actually stands for. In face of the various
approximation schemes that are involved such as the large $s$
expansion, the continuum limit etc. it is not always clear which
features of the quantum spin chain are captured by continuum field
theory and which are due to, say, the lattice.

The principal objective of this paper is to revisit the Haldane
mapping of the DSC and investigate whether it can be used, like
the large $N$ expansion of the $CP^{N-1}$ model, as an explicit
example for demonstrating the super universal features of
the instanton vacuum. We are interested, in particular, in
the specific mechanism that is responsible for the
generation of {\em robust} topological quantum numbers, i.e. the
QHE itself.

In order to achieve our goals we shall begin in Section II with a
review of edge excitations in the Grassmannian NLSM as well as the
simple example of a single Heisenberg spin in a magnetic field
$B$. From the {\em path integral representation} (PIR) we readily
establish the fact that the dynamics of the isolated spin is
exactly described by the $\theta$ term (solid angle term) that one
normally would associate with the $O(3)$ NLSM. The time
correlations that we obtain precisely correspond to those obtained
earlier, describing the {\em massless chiral edge modes} in
quantum Hall systems. This clearly indicates that the {\em
dangling spins} at the edge of the chain, just like the {\em edge
currents} in quantum Hall systems, are responsible for the low
energy dynamics of the instanton vacuum in strong coupling. It is
important to remark, though, that our phrase ``edge correlations''
is not obviously related to the {\em chiral edge boson} approach
in $1+1$ space-time dimensions that one usually associates with
the QHE. Although one deals with essentially the same physical
phenomenon, the formal equivalence between the two theories is by
no means any obvious and the matter has been addressed in great
detail in Refs ~\cite{Unify3} and ~\cite{Unify4}.

The results for the single spin motivate us to proceed and
investigate the Haldane mapping for spin chains with and without
an edge in Section III. We consider the closed chain, open chains
with an even/odd number of spins as well as the half infinite
chain and establish their {\em topological differences}.  These
differences are reflected by the fact that for each case the
$\theta$ parameter in the continuum field theory has a distinctly
different physical meaning. For specificity we shall associate
{\em topologically} distinct quantum Hall systems with each of the
{\em geometrically} different spin chain in Section IV.

Armed with these insights we next address the various strong
coupling aspects of the Haldane mapping in greater detail.  We
focus the attention primarily on the problem of {\em weakly
coupled dimers} where one naively would expect that the Haldane
mapping can no longer be trusted. To deal with this problem in all
its generality we set up in Section V an {\em exact}
renormalization group decimation scheme that can be applied
directly to the PIR of the DSC. Here, the phrase {\em exact} means
that in the limit of completely decoupled dimers the
renormalization group equations can be systematically expanded order
by order in the small dimerization parameter $\kappa$. The
important advantage of this scheme is that it can be applied to
all geometrically different spin chains under consideration ({\em
open} and {\em closed} spin chains etc.).

Next we specify to the $s=1/2$ case and obtain explicit results
from the decimation procedure. These reveal several novel and
important aspects of the problem that ultimately lay the bridge
between the DSC and continuum field theory.

\begin{enumerate}

\item
First of all, the renormalization group takes us away from the
pure DSC and the system ``flows" into a more general dimerized
chain, termed the fermionic rotor chain (FRC). Unlike the DSC, the
FRC involves in total four different coupling constants $\kappa_1,
.. \kappa_4$ rather than the dimerization parameter $\kappa$
alone.
\item
The renormalization group equations can be solved exactly, the
solution indicating that the FRC has a {\em massive} fixed point
as expected.  However, the mass gap of the FRC is actually a
highly nonperturbative expression in the coupling constants.
\item
The decimation procedure {\em dynamically} generates a {\em dangling
spin} with $s=1/2$ at the edge of the chain.  The renormalization
group procedure demonstrates explicitly that the spin at the edge
becomes completely decoupled from the bulk only after a sufficiently
large number of iterations.
\item
Guided by the results of the decimation procedure we next derive
in Section VI the NLSM in the presence of the $\theta$ term,
describing the low energy dynamics of the FRC. The derivation
proceeds analogous to the Haldane mapping. We make use of the fact
that the two rotors on each dimer are almost anti-parallel. The
deviation from being anti-parallel are then the ``hard" fields
which are integrated out.  As the final step we perform the
continuum limit which then leads to explicit expressions of the
NLSM parameters (the coupling constant $g$ and the instanton angle
$\theta$) in terms of the FRC parameters $\kappa_1,.... \kappa_4$.
\item
At this stage of the analysis all the results that we have
obtained become simultaneously important. First of all the
renormalization group equations of the FRC directly imply that the
NLSM parameters $g$ and $\theta$ are {\em running} parameters with
length scale. Hence, our results explicitly display the important
feature of {\em $\theta$ renormalization}.  Secondly, the
effective NLSM action associated with the ``massive" fixed points
of the FRC precisely describes the time correlations of the {\em
dangling spin} at the edge or, equivalently, {\em massless chiral
edge excitations} in quantum Hall systems. Thirdly, the general
renormalization group scenario that emerges, in the inverse
coupling constant $1/g$ versus $\theta$ plane, explicitly
demonstrates how the QHE emerges as a super universal strong
coupling feature of the instanton vacuum. We identify the
Hall conductance in the system and show that it is robustly
quantized with corrections that are exponentially small in the
size of the chain. We end this paper with a conclusion in Section
VII.

\end{enumerate}
\section{Physics at the edge}
\label{epsc}

\subsection{The strong coupling limit}
\label{bscl}

Within the replica field theory approach to localization problems
the low energy dynamics of quantum Hall regime is described by the
Grassmannian $U(N+M)/U(N)\times U(M)$ NLSM. The effective action
is defined as follows~\cite{Prange}
\begin{eqnarray}
\nonumber
 S[Q]&=&-\frac{1}{8}\sigma_{xx}^0\int d^2x {\rm tr}(\nabla Q)^2\\
 \nonumber &&+\frac{1}{8}\sigma^0_{xy}\int d^2x~{\rm
 tr}~\epsilon_{ij}Q
 \partial_i Q\partial_j Q\\
 &&+\pi\rho_0\omega \int d^2x {\rm tr} \Lambda Q.
 \label{Ssigma}
\end{eqnarray}
The field variable $Q(x)$ can be represented according to
\begin{equation}
 Q=T^{-1}\Lambda T,~~ T\in U(N+M)
\end{equation}
where $\Lambda$ is a diagonal matrix with $N$ diagonal elements
equal to $1$ and the other $M$ equal to $-1$
\begin{equation}
 \Lambda = \left(
 \begin{matrix}
 \mathbf{1}_N & 0 \\ 0 & -\mathbf{1}_M
 \end{matrix}
 \right) .
\end{equation}
The results for the electron gas are obtained by taking the {\em
replica limit} $N,M\rightarrow 0$ at the end of all computations.
As mentioned earlier, the coupling constants $\sigma_{xx}^0$ and
$\sigma_{xy}^0$ are dimensionless quantities representing the mean
field  {\em longitudinal} and {\em Hall} conductances
respectively. The quantity $\rho_0$ is the density of electronic
levels and $\omega$ is the external frequency that is normally
used to regulate the infrared of the system.

To study the theory in the DC limit ($\omega =0$) we have to
specify the boundary conditions satisfied by the $Q$ fields. As
always, this has to be decided by the physics of the system under
consideration. It was shown
~\cite{PruiskenBaranovVoropaev,PruiskenBurmistrov,Unify3,Unify4}
that ``free" boundary conditions (i.e. no boundary conditions on
the matrix field variables $Q$) implies the existence of {\em
massless chiral edge excitations}. In this Section we briefly
review these ideas and show that they resolve some of outstanding
strong coupling problems associated with the topological concept
of an instanton
vacuum.~\cite{PruiskenBaranovVoropaev,PruiskenBurmistrov}
\subsubsection{Fractional topological charge and chiral edge excitations}
The most important consequence of free boundary conditions is that
the topological charge $\mathcal{C} [Q] $ of the theory is not
quantized. By writing an arbitrary matrix field $Q$ as follows,
\begin{equation}
\label{edbu} Q=t^{-1}Q_0 t,
\end{equation}
where $Q_0$ has a fixed value at the edge, say $\Lambda$, and $t$
represents the fluctuations about these special boundary
conditions. Then in general we can write,
\begin{eqnarray}
 \mathcal{C} [Q]&\equiv&\frac{1}{16\pi i}\int d^2x~{\rm tr}~
 \epsilon_{ij}Q\partial_i Q\partial_j Q \\
 &=& \mathcal{C} [Q_0 ] + \mathcal{C} [q] .
 \label{ftopch}
\end{eqnarray}
Here, $\mathcal{C} [Q_0 ]$ is by construction {\em integer}
quantized. The quantity $\mathcal{C} [q]$ with $q=t^{-1} \Lambda
t$ represents the {\em fractional} part and without loss of
generality we can write $-1/2 \leq \mathcal{C} [q] <1/2$.

The fundamental significance of the fractional charge $\mathcal{C}
[q]$ becomes particularly transparent if we consider the special
case where the Fermi energy of the electron gas lies in a Landau
gap. Examining this physical situation is important since it
corresponds to the naive strong coupling limit of the theory
obtained by putting $\sigma_{xx}^0 = \rho_0 = 0$ and
$\sigma_{xy}^0 = k$ with the integer $k$ denoting the number of
completely filled Landau bands. The action becomes simply
\begin{equation}
    S[Q] = \frac{k}{8}\int d^2x~{\rm tr}~
\epsilon_{ij}Q\partial_i Q\partial_j Q
\label{Sgap}
\end{equation}
which can be written as
\begin{equation}
    S[Q]=2\pi ik~\mathcal{C} [Q_0]
    +\frac{k}{2}\oint d\vec x\cdot {\rm tr}(\Lambda t\nabla t^{-1})
\label{Ssplit}
\end{equation}
The significance of the integer valued Hall conductance
$\sigma_{xy}^0=k$ is now clear. In particular, since
$\sigma^0_{xx}=0$ the theory renders completely insensitive to the
field configurations $Q_0$ that have an {\em integer quantized}
topological charge. These configuration are generally identified
as the {\em bulk excitations} of the theory. The action solely
depends only on the fluctuating matrix fields $t$ along the one
dimensional edge of the system. Hence, one can generally identify
the {\em edge excitations} of the system with those matrix field
configurations that carry a {\em fractional} topological charge.

It turns out that the complete expression for the effective action
along the edge is given
by~\cite{PruiskenBaranovVoropaev,PruiskenBurmistrov}
\begin{equation}
 S[t]=\frac{k}{2}\oint d\vec x\cdot {\rm tr}(\Lambda t\nabla
 t^{-1}) + \omega \rho_{edge} {\rm tr}\Lambda q .
 \label{fulledge}
\end{equation}
Here, the quantity $\rho_{edge}$ indicates that although the
density of levels is zero in the {\em bulk} there nevertheless
exists a finite density of electronic levels at the {\em edge} of
the system that can carry the Hall current.
\subsubsection{Chiral edge fermions and chiral edge bosons}
It should be emphasized that the action for the {\em edge}
(Eq.~\ref{fulledge}) was originally obtained as a special case of
a {\em bulk} theory that does not involve any explicit knowledge
on the microscopic details of the edge of the electron
gas.~\cite{Prange} It can remarkably be shown, however, that
Eq.~\eqref{fulledge} is completely equivalent to a system of
disordered {\em chiral fermions} that are confined at the edge of
the system.~\cite{PruiskenBaranovVoropaev} The replicated chiral
fermion action $S_{CF}$ is given in terms of $k$ independent edge
modes according to
\begin{eqnarray}
 \label{rfact}
 S_{CF} (V) &=& \int \sum_{jj^\prime =1}^k {\rm tr}
 \Psi_{j}^\dag \left[ -iv_d\partial_x  + V^{jj^\prime} +i\omega \Lambda
 \right] \Psi_{j^\prime} .~~~~
\end{eqnarray}
Here, $\Psi_{j}$ and $ \Psi_{j}^\dag$ represent $(N+M)$-component
vectors of fermion fields which can be written as $\left[
\psi_{j\alpha}^{(+)} (x), \psi_{j\beta}^{(-)} (x) \right]$ and
$\left[ \bar{\psi}_{j\alpha}^{(+)} (x), \bar{\psi}_{j\beta}^{(-)}
(x) \right]^\dag$ respectively with $\alpha =1,2, ...N$ and
$\beta=1,2, ...M$. The hermitian matrices $V^{jj^\prime} (x)$ are
randomly distributed potentials (with a Gaussian weight) that
facilitate scattering between the edge channels labeled by $j$
and $j^\prime$.

The equivalence of Eqs~\eqref{rfact} and ~\eqref{fulledge} can
formally be demonstrated on the basis of the standard techniques
using the Hubbard-Stratonovich transformation. However, an
important aspect of chiral electrons is that the quenched
randomness such as $V^{jj^\prime}$ can be ``gauged away". This
means that the ``gauge invariant" correlations of the system can
be computed in a trivial and exact manner simply by putting the
random matrices $V^{jj^\prime}$ in Eq.~\eqref{rfact} equal to
zero. Several important examples are as
follows~\cite{PruiskenBaranovVoropaev}
\begin{equation}\label{OneQ}
\langle q^{pp^\prime}_{\alpha\beta} (x) \rangle_q =
\rho_{edge}^{-1} \sum_{j=1}^k \langle {\bar\psi}^p_{j
\alpha}(x)\psi^{p^\prime}_{j \beta}(x) \rangle_{CF} = \left(
\Lambda \right)^{pp^\prime}_{\alpha\beta} .
\end{equation}
Here, the expectations $< ... >_q$ and $< ... >_{CF}$ are with
respect to the theory of Eq.~\eqref{fulledge} and
Eq.~\eqref{rfact} with $V^{jj^\prime} = 0$ respectively.
Furthermore, we have made use of the following relation between
the density of edge levels $\rho_{edge}$ and the drift velocity
$v_d$ of the chiral electrons
\begin{equation}
\rho_{edge} = m/v_d .
\end{equation}
Eq.~\eqref{OneQ} surprisingly shows that the continuous
$U(N+M)/U(N) \times U(M)$ symmetry is {\em permanently broken} at
the edge of the system, independent of the values of $N$ and $M$.
The higher order correlation functions can be computed along
similar lines. The only non-vanishing two point function is given
by~\cite{PruiskenBaranovVoropaev}
\begin{eqnarray}
 && \langle q^{-+}_{\alpha \beta}(x) q^{+-}_{\alpha
 \beta}(y)\rangle_q  \nonumber \\
 && =\rho_{edge}^{-2} \sum_{jj^\prime =1}^k \langle {\bar\psi}^{(-)}_{j
 \alpha}(x)\psi^{(+)}_{j \beta}(x) {\bar\psi}^{(+)}_{j^\prime
 \alpha}(y)\psi^{(-)}_{j^\prime \beta}(y) \rangle_{CF} ~~~~~~\\
 && = k  \vartheta(x-y)e^{-2\omega \rho_{edge} (x-y)} \label{2cf}
\end{eqnarray}
for $x \neq y$ and $\vartheta$ denoting the Heaviside step
function. This result clearly shows that in the limit $\omega = 0$
the theory is length scale independent and, hence, our one
dimensional edge theory is a {\em critical} theory.

In summary we can say that the instanton vacuum displays {\em
massless chiral edge excitations}. This remarkable and exactly
solvable feature is independent of the number of field components
(replica's) $N$ and $M$ in the theory. We have focused so far on
the ``naive" strong coupling limit of the theory which in the
language of the electron gas corresponds to the very special case
of integer filling fractions $\nu = k$ where the system displays
an energy gap or {\em Landau gap}. In the next Section we will
show, however, that the result of Eq.~\eqref{fulledge} actually
has a much broader range of validity and quite generally emerges
as the {\em critical fixed point action} of the quantum Hall
state. Although experimentally well established, the quantum Hall
state says something highly non-trivial and remarkably general
about the strong coupling features of the instanton vacuum concept
that previously have remained concealed.
\subsubsection{Quantization of the Hall conductance}
To address the theory in all its generality it is necessary to
find a way in which the massless edge modes $t$ are unambiguously
separated from the distinctly different bulk matrix field
variables $Q_0$. This separation becomes immediately transparent
once it is recognized that the mean field quantity $\sigma_{xy}^0$
can generally be expressed as the sum of an {\em integer
quantized} ``edge" piece $k$ and an {\em unquantized} ``bulk"
piece $\theta_{bulk}$. Working for simplicity in the limit of
strong magnetic fields $B$ then $\sigma_{xy}^0$ is identical to
the filling fraction~\cite{Prange} $\nu$ such that we can write
\begin{eqnarray}
 \sigma^0_{xy} \equiv \nu = k(\nu) + \theta_{bulk} (\nu)/2\pi
 .\label{sigmaxysplit}
\end{eqnarray}
where
\begin{eqnarray}
 k(\nu) \in \mathbb{Z}, ~~ -\pi \leq \theta_{bulk} (\nu) < \pi
\end{eqnarray}
\begin{figure}
\centerline{\includegraphics[width=.55\textwidth]{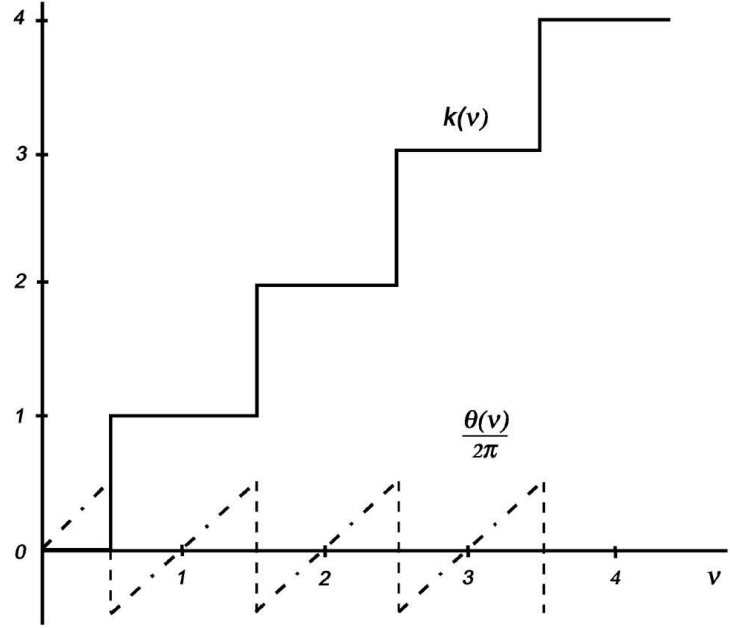}} \caption{The
quantity $\sigma_{xy}^0 = \nu$ is the sum of a {\em quantized}
edge part $k(\nu)$ and an {\em unquantized} bulk part
$\theta(\nu)$ .}\label{thetanu}
\end{figure}
In Fig.~\ref{thetanu} we sketch the $k(\nu)$ and $\theta_{bulk} (\nu)$
with varying $\nu$. To see how the split in
Eq.~\eqref{sigmaxysplit} facilitates a general discussion of the
theory on the strong coupling side we first consider the
topological piece of the action which now can be written as
follows
\begin{eqnarray}
 && \frac{1}{8}\sigma^0_{xy}\int ~{\rm tr}\epsilon_{ij} Q
 \partial_i Q\partial_j Q =  2\pi i \sigma_{xy}^0 \mathcal{C} [Q]
 \nonumber \\ \nonumber \\
 &&~~~~~~~ =  2\pi i \left\{ k(\nu) + \theta_{bulk} (\nu)/2\pi \right\} \left\{
 \mathcal{C} [Q_0] + \mathcal{C} [q] \right\} ~~~~~~\nonumber \\ \nonumber \\
 &&~~~~~~~ = 2\pi i k(\nu) \mathcal{C} [q] + i \theta_{bulk} (\nu) \mathcal{C}
 [Q].
\end{eqnarray}
In the last step we have left out the term $2\pi i k(\nu)
\mathcal{C} [Q_0]$ since it gives rise to phase factors which are
not of interest to us. On the basis of this result we can split
the action $S[Q]$ into an edge piece and a bulk piece as follows
\begin{eqnarray}
 S[Q]&=& S_{edge} [q] + S_{bulk} [Q] \\
\label{edscac}
 S_{edge} [q] &=& 2\pi i k(\nu) \mathcal{C} [q] = \frac{k(\nu)}{2} \oint
{\rm tr}
 t \partial_x t^{-1} \Lambda \\
 S_{bulk} [Q] &=& -\frac{1}{8}\sigma_{xx}^0 (\nu) \int {\rm tr}(\nabla Q)^2
 + i \theta_{bulk} (\nu) \mathcal{C}
 [Q]. ~~~~~\label{Ssigma1}
\end{eqnarray}
Notice that in the very special case of an integer valued filling
fraction $\nu$ the bulk part of the action $S_{bulk}$ is trivially
equal to zero and we recover the results of the previous Section.
On the other hand, to deal with the general situation of arbitrary
$\nu$ we must perform the integration over the bulk matrix field
variables $Q_0$. Provided the edge matrix field variables $q$ obey
the classical equations of motion in the bulk of the system one
can express the results in terms of an effective action
\begin{equation}
 \label{edeff}
 e^{S_{bulk}^\prime [q]}=\int_{\partial V} D[Q_0] e^{S_{bulk}
 [t^{-1}Q_0t]}.
\end{equation}
The subscript ${\partial V}$ reminds us of the fact that the
functional integral has to be performed with a fixed value $Q_0 =
\Lambda$ at the edge. Next, on the basis of very general symmetry
considerations ~\cite{PruiskenBaranovVoropaev} one concludes that
the effective action $S_{bulk}^\prime [q]$ has to be of the form
\begin{eqnarray}
 S_{bulk}^\prime [q] = -\frac{1}{8}\sigma^\prime_{xx}
 \int {\rm tr}(\nabla q )^2 +i\theta_{bulk}^\prime
 \mathcal{C} [q].
 \label{Seff}
\end{eqnarray}
The total effective action for the edge field variable $q$ becomes
\begin{eqnarray}
 S_{eff} [q] &=& S_{edge}^\prime [q] + S_{bulk}^\prime [q] \\
 &=& -\frac{1}{8}\sigma^\prime_{xx}
 \int {\rm tr}(\nabla q )^2 + 2\pi i ( k(\nu)+\theta_{bulk}^\prime
 /2\pi) \mathcal{C} [q].\nonumber \\
 \label{Seff1}
\end{eqnarray}
Here, the quantities $\sigma_{xx}^\prime$ and $\sigma_{xy}^\prime
= k(\nu) + \theta_{bulk}^\prime /2\pi$ are identified as the
macroscopic transport coefficients of the system, i.e. the Kubo
formulae for the longitudinal and Hall conductance
respectively.\cite{Prange}

In spite of the fact that Eqs~\eqref{edeff} and ~\eqref{Seff} are
extremely complicated expressions as they stand, they nevertheless
provide a profound and general framework that is needed for a
microscopic understanding of the quantum Hall effect. For example,
since the non-linear $\sigma$ model is known to be asymptotically
free in two dimensions for all non-negative integer values of $N$
and $M$, one generally expects the theory to {\em dynamically}
develop a mass gap in the limit of large distances. On the other
hand, from the definition of $\sigma_{xx}^\prime$ and
$\theta_{bulk}^\prime$ we clearly see that these quantities play
the role of {\em response parameters} that measure the sensitivity
of the bulk of the system to an infinitesimal change in the
boundary conditions. These quantities should therefore render
exponentially small in the system size $L$ provided the bulk of
the system displays a mass gap or a finite correlation length
(localization length) $\xi$. Putting the various different
statements together we obtain the following general expressions
for the quantum Hall effect
\begin{eqnarray}
 \sigma^\prime_{xx} &=& \mathcal{O} (e^{-L/\xi}) \\
 \sigma^\prime_{xy} &=& k(\nu) + \theta_{bulk}^\prime /2\pi
 \nonumber \\
 &=& k(\nu) + \mathcal{O} (e^{-L/\xi}) .
 \label{qHe}
\end{eqnarray}
It now is clear why the critical edge excitations play such a
fundamental role in this problem. Following Eq.~\eqref{Seff1} we
see that the effective action of the quantum Hall state precisely
equals the action for massless chiral edge excitations
$S_{edge}$. This explains why the quantity $k(\nu)$ entering the
expression for the Hall conductance (Eq.~\ref{qHe}) is {\em length
scale independent} and, hence, {\em quantized}. This is quite
unlike the bulk part of the Hall conductance $\theta_{bulk}^\prime
/2\pi$ which probes the mass gap in the bulk of the system. This
quantity therefore determines the {\em corrections} to exact
quantization which, as mentioned above, are exponentially small in
the system size.

Notice that nothing much of the argument seems to depend on the
specific application of the instanton vacuum that one considers.
One therefore expects that the theory generically displays the
principal aspects of the quantum Hall effect, independent of the
values of $N$ and $M$. It is important to emphasize, however, that
none of these general strong coupling features of the theory have
previously been recognized, in particular the fundamental
significance of $\theta$ {\em renormalization}. It is therefore
extremely important to have certain explicit examples of an
instanton vacuum where the statements of Eq.~\eqref{qHe} can be
explored and investigated in detail. The rest of this paper will
be devoted to developing a formalism that enables us to
analytically access the strong coupling quantum Hall fixed points.

We end this Section with several remarks. First of all, it should
be mentioned that the statement of Eq. ~\eqref{sigmaxysplit} has a
quite general significance which is not limited to strong coupling
phenomena alone. In fact, both the motivation and justification of
the split in $\sigma_{xy}^0$ is provided by the detailed knowledge
obtained from the ``observable" parameters $\sigma_{xx}^\prime$
and $\theta_{bulk}^\prime$ in the weak coupling regime. The
results can generally be expressed in terms of renormalization
group $\beta$ functions according to ~\cite{PruiskenBurmistrov}
\begin{eqnarray}
\sigma_{xx}^\prime &=& \sigma_{xx}^0 +
\int_{\lambda_0}^{\lambda^\prime} \frac{d\lambda}{\lambda}
\beta_\sigma ( \sigma_{xx} , \theta_{bulk}) \\
\theta_{bulk}^\prime &=& \theta_{bulk}^0 +
\int_{\lambda_0}^{\lambda^\prime} \frac{d\lambda}{\lambda}
\beta_\theta ( \sigma_{xx} , \theta_{bulk}).
\end{eqnarray}
Here, the $\lambda^\prime$ and $\lambda_0$ denote the length
scales that are generally associated with the observable theory
$\sigma_{xx}^\prime , \theta_{bulk}^\prime$ and bare theory
$\sigma_{xx}^0 , \theta_{bulk}^0 = \theta_{bulk} (\nu)$
respectively. By using otherwise very general statements of
symmetry one can show that the $\beta$ functions can be written in
terms of an infinite trigonometric series in discrete topological
sectors of the theory according to
\begin{eqnarray}
 \beta_\sigma &=& \frac{d\sigma_{xx}}{d\ln\lambda} =
 \sum_{n=0}^\infty f_n (\sigma_{xx}) \cos
 n\theta_{bulk} \\
 \beta_\theta &=& \frac{d\theta_{bulk}}{d\ln\lambda}
 =\sum_{n=1}^\infty g_n (\sigma_{xx}) \sin n\theta_{bulk}
\end{eqnarray}
These results tell us that the lines $\theta_{bulk} = \pm \pi$ and
$\theta_{bulk} = 0$ are invariant under the action of the
renormalization group. On the other hand, an explicit computation
of the lowest order terms in the series leads to the following
results~\cite{PruiskenBurmistrov}
\begin{eqnarray}
 f_0 (\sigma_{xx}) &=&  -(N+M) - \frac{NM +N+M}{\sigma_{xx}} +
 \mathcal{O}(\sigma_{xx}^{-2}) ~~~~~\\
 f_1 (\sigma_{xx}) &=& 2\pi g_1 (\sigma_{xx})
 = - D_{N,M} \sigma_{xx} e^{-2\pi\sigma_{xx}}
\end{eqnarray}
where the positive quantity $D_{N,M}$ is determined by the
instanton determinant.~\cite{PruiskenBurmistrov} These results
clearly indicate that the renormalization group flow is toward the
strong coupling fixed point $\sigma_{xx} = \theta_{bulk} =0$ which
is stable in the infrared.

Notice that the $\beta$ functions can also be written in terms of
the Hall conductance $\sigma_{xy}$ simply by replacing
$\theta_{bulk} /2\pi$ by $k(\nu) + \theta_{bulk} /2\pi$. This
leads to the well known statement which says that the
renormalization group flow is {\em periodic} in $\sigma_{xy}$.
This statement, however, is equivalent to the following statement
which says that
\begin{equation}
\frac{dk((\nu)}{d\ln\lambda} = 0. \label{critedge}
\end{equation}
As we have seen earlier, the true justification of this result
and, hence, of the aforementioned {\em periodicity} statement in
the $\sigma_{xx}$, $\sigma_{xy}$ conductance plane lies in the
{\em critical} behavior of the edge of the instanton vacuum. This
changes the meaning of Eq.~\eqref{critedge} into that of a {\em
critical fixed point}.

\subsection{Edge spins}
\label{espins}

For the rest of this paper, we will focus on the $N=M=1$ case. The group,
$SU(N+M)$ is then $SU(2)$ and the Grassmannian is the 2-sphere.
$Q$ parameterizes the points on a 2-sphere and we can write it as,
\begin{equation}
Q={\hat n}\cdot{\vec \tau}
\end{equation}
Where $\hat n$ is a unit vector. The fractional part of the $\theta$ term
is the solid angle of the curve traced out on the sphere by $\hat n$ as it
goes around the edge. The action (\ref{Sgap}) can be written as,
\begin{equation}
\label{spinac}
S[Q]=is\int d^2x~{\hat n}\cdot\partial_x{\hat n}\times \partial_y{\hat n}
+2\pi\rho_{edge}\omega n^3
\end{equation}
with $s=\frac{m}{2}$. This is exactly the action obtained for the
problem of a spin-$s$ system \cite{perel, fradkin}, if the
Euclidean time of the spin system is identified with the
coordinate that parameterizes the edge, say $x$. The other
coordinate, $y$, that goes into the bulk plays the role of the
fictitious dimension that it is necessary to introduce in the spin
problem in order to write down a globally well defined action.

The path integral for spin systems is standardly derived using spin-coherent
states \cite{perel, fradkin}.  The spin-$s$ system is an
irreducible representation of the angular momentum algebra,
\begin{eqnarray}
\label{angmom}
\left[J^a,J^b\right]&=&i\epsilon^{abc}J^c\\
\nonumber
J^aJ^a\vert \psi \rangle &=& s(s+1)\vert \psi \rangle
\end{eqnarray}
The states can be labeled by the eigenvalue of $J^3$,
\begin{equation}
\label{sphstates}
J^3\vert M \rangle = M \vert M \rangle,~~~M=-s,...,s
\end{equation}
The spin coherent states are defined as,
\begin{equation}
\label{spcoh}
\vert Q \rangle = T\vert -s \rangle
\end{equation}
Where, $T\in SU(2)$. The little group of the state $\vert -s
\rangle$ is $U(1)$. So the states are in one-to-one correspondence
with points on the 2-sphere. They are labeled by $Q =
-T\tau^3T^{-1}$. If we take the Hamilton to be,
\begin{equation}
\label{spham}
H= 4\pi\frac{\rho_{edge}}{2s}J^3
\end{equation}
Then the path integral representation for the partition function is,
\begin{eqnarray}
Z&=&{\rm tr}e^{-LH}\\
&=&\int {\cal D}\left[Q(x)\right]~e^{-S\left[Q(x)\right]}
\end{eqnarray}
Where $dQ$ denotes a normalized measure on the two sphere and the
action is given by Eq.~(\ref{spinac}). Hence the edge theory
defined by the action in Eq.~(\ref{Sgap}), is exactly equivalent
to a spin-$\frac{m}{2}$ system with the Hamiltonian given by
(\ref{spham}). The correlation functions of the $Q$ fields can be
expressed in the operator formalism as,
\begin{eqnarray}
\nonumber
s^n\langle Q_{p_1p^\prime_1}(x_1)...Q_{p_np^\prime_n}(x_n)\rangle
&=& {\rm tr} \Big( e^{-LH}{\cal T}_x
\big( J_{p_1p^\prime_1}(x_1)... \\
&&...J_{p_np^\prime_n}(x_n) \big) \Big),
\end{eqnarray}
where $J_{pp^\prime}\equiv J^a\tau^a_{pp^\prime}$. The (euclidean)
equations of motion are easily solved,
\begin{eqnarray}
J^3(x)&=&J^3(0)\\
J^+(x)&=&J^+(0)e^{\frac{\omega}{v_d}x}\\
J^-(x)&=&J^-(0)e^{-\frac{\omega}{v_d}x}
\end{eqnarray}
The only non-zero two point function at $L=\infty$ is easily computed
to be,
\begin{equation}
\langle Q^{-+}(x_1)Q^{+-}(x_2)\rangle
=m\theta(x_1-x_2)e^{-2\frac{\omega}{v_d}(x_1-x_2)}
\end{equation}
which is exactly the result in Eq.(\ref{2cf}). Thus we have shown
that the critical edge action and correlation functions obtained
in the bare strong coupling limit is exactly the same as that of a
single spin-$\frac{m}{2}$ system for $N=M=1$. These results can be
generalized for all $N$ and $M$. We will be presenting the results
in a forthcoming publication.


\section{Dimerized spin chains}
\label{dsc}

\begin{figure}
\psfrag{a}[l,b][l,b]{$a)$}
\psfrag{b}[l,b][l,b] {$b)$}
\psfrag{c}[l,b][l,b]{$c)$}
\psfrag{j}[l,b][l,b]{$J$}
\psfrag{jt}[l,b][l,b]{$\tilde{J}$}
\psfrag{B}[l,b][l,b]{${\bf B}$}
\psfrag{k}[l,b][l,b]{$\kappa$}
\psfrag{t}[l,b][l,b]{$\tilde{\kappa}$}
\psfrag{i1}[c][c]{$I=1$}
\psfrag{i2}[c][c]{$I=2$}
\psfrag{i3}[c][c]{$I=3$}
\psfrag{1}[c,t][c,t]{$1$}
\psfrag{2}[c,t][c,t]{$2$}
\psfrag{E}[c,t][c,t]{$E$}
\begin{center}
\includegraphics[width= .55\textwidth]{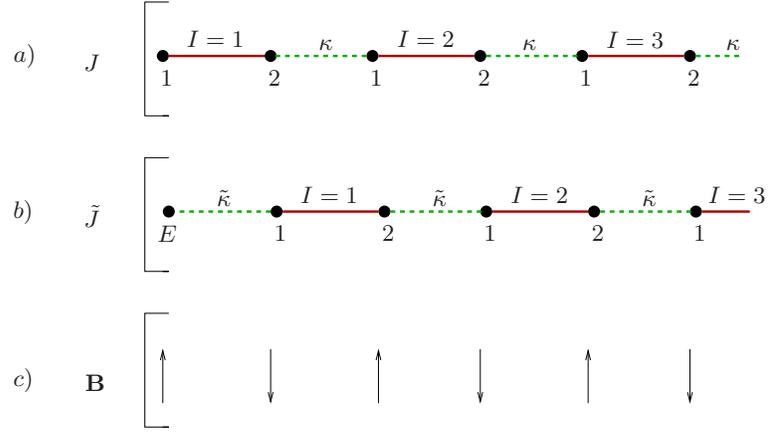}
\end{center}
\caption{\label{dimer}Semi-infinite dimerized spin chain with no
  edge spin $(a)$ and with the edge spin in the dual representation
  $(b)$. The orientation of the staggered magnetic field is shown in $(c)$.}
\end{figure}
The results of the previous Section implies that the bare strong
coupling limit of the NLSM is exactly equivalent to the low energy
physics of spin-$\frac{m}{2}$ dimerized spin chain (DSC), in the
strong dimerization limit. We first consider the case of
semi-infinite DSC's and will discuss the case of finite chains
later. The Hamiltonian is,
\begin{equation}
\label{dcham}
H_{DSC}= \sum_{I=0}^\infty J\left( {\bf S}_{2i}\cdot{\bf S}_{2i+1}
+\kappa {\bf S}_{2i+1}\cdot{\bf S}_{2i+2}\right).
\end{equation}
At $\kappa=0$, the model consists of decoupled dimers and is trivially
solved. All dimers being in the singlet state constitutes the ground
state. Any one of the dimers being in the triplet state constitute the
lowest energy excitations. Thus the system has a gap equal to $J$ and
there are no excitations at energy scales small compared to $J$.

At $\kappa=\infty$ also, the model decouples. The Hamiltonian can be written
in a {\em dual} representation where the roles of the `dimers' and the `weak
bonds' get interchanged (see Fig.~\ref{dimer}),
\begin{equation}
{\tilde H}_{DSC}= \sum_{i=0}^\infty {\tilde J}\left(
{\tilde \kappa}~{\bf S}_{2i}\cdot{\bf S}_{2i+1}
+{\bf S}_{2i+1}\cdot{\bf S}_{2i+2}\right).
\end{equation}
where we have defined ${\tilde \kappa}\equiv \frac{1}{\kappa},~
{\tilde J}\equiv \kappa J$. Thus
$\kappa=\infty\Leftrightarrow{\tilde \kappa}=0$. The system
now decouples into a set of dimers and a free spin at the edge. The
dynamics at energy scales small compared to $J$ will therefore
be that of the free spin at the edge (see Fig.~\ref{dimer}b).

Thus, consistent with the Haldane mapping, the low energy physics of the
DSC at $\kappa=0$ and $\kappa=\infty$ is the same as that of the NLSM at
$\sigma_{xx}=0=\sigma_{xy}$ and $\sigma_{xx}=0,~\sigma_{xy}=m$ respectively.
In the former case both models have no low energy dynamics and in the latter
case both have low energy dynamics confined to edge with identical correlation
functions.

\subsection{Haldane mapping with an edge}
\label{hm}

The above validation of the Haldane mapping in the strong coupling limit is of
course somewhat trivial at this stage. We will later be developing formalism
to justify the mapping at non-zero values of $\kappa (\tilde \kappa)$. In
this section we will perform the Haldane mapping, with the standard
approximations, of the semi-infinite DSC to the NLSM with an edge.
We point out some features and also discuss the approximations involved.

For $\kappa \le 1$, it is useful to change the labeling of the spins and
denote,
\begin{eqnarray}
\nonumber
{\bf S}_{I,1}&\equiv&{\bf S}_{2I},\\
{\bf S}_{I,2}&\equiv&{\bf S}_{2I+1}.
\end{eqnarray}
The Hamiltonian is then written as,
\begin{equation}
H= J \left[ \sum_{I=0}^\infty {\bf S}_{I,1} \cdot {\bf S}_{I,2}
+ \kappa  \sum_{I=0}^\infty {\bf S}_{I,2} \cdot {\bf S}_{(I+1),1}
\right].
\end{equation}
For $\kappa\ge1$, we denote,
\begin{eqnarray}
\nonumber
{\bf S}_E&\equiv&{\bf S}_0,\\
\nonumber
{\bf S}_{I,1}&\equiv&{\bf S}_{2I+1},\\
{\bf S}_{I,2}&\equiv&{\bf S}_{2I+2}.
\end{eqnarray}
The Hamiltonian in the dual representation is then,
\begin{equation}
\tilde{H} =  \tilde{J} \left[ { \bf S}_E \cdot { \bf S}_{0,1}
+ \tilde{\kappa} \sum_{I=0}^\infty {\bf S}_{I,1} \cdot {\bf S}_{I,2}
+ \sum_{I=0}^\infty {\bf S}_{I,2} \cdot {\bf S}_{I+1,1} \right].
\end{equation}

The Hamiltonians $H$ and $\tilde{H}$ indicate that apart from edge
effects the spin system has a {\em dual symmetry}

\begin{eqnarray}
\kappa & \rightarrow & \kappa^{-1},
\nonumber \\
J & \rightarrow & \kappa J.
\end{eqnarray}
This symmetry has a different meaning dependent on the value $s$ of
the spin. In what follows we shall separately derive the effective
action of the spin system in the completely equivalent representations
given by $H$ and $\tilde{H}$ respectively.

To proceed it is helpful to introduce a staggered magnetic field
$\bf B$ in $H$.
\begin{equation}
H \rightarrow  H + \sum_{I=0}^\infty {\bf B} \cdot ({\bf S}_{I,1}
- {\bf S}_{I,2}).
\end{equation}
This term favours an anti-ferromagnetic spin arrangement of the
semi-infinite chain (see Fig.~\ref{RGflowMN}c). Notice that the same
anti-ferromagnetic order is induced by adding the following terms
to $\tilde{H}$.
\begin{equation}
\tilde{H} \rightarrow  \tilde{H} + {\bf B_0} \cdot { \bf S}_E
- \sum_{I=0}^\infty {\bf B} \cdot ({\bf S}_{I,1} - {\bf S}_{I,2}).
\end{equation}
Next it is convenient to introduce a slightly different notation
for the dimer terms in the theory. Without loss of generality we may
replace the terms ${\bf S}_{I,1} \cdot {\bf S}_{I,2}$ in $H$ and
$\tilde{H}$ by the expression $\frac{1}{2}({\bf S}_{I,1} +
{\bf S}_{I,2})^2$. Keeping this in mind we obtain the euclidean
action in the coherent state basis as follows.
\begin{equation}
S = is \sum_{I=0}^\infty  \Omega [ \hat{{\bf n}}_{I,1} ] +
\Omega [ \hat{{\bf n}}_{I,2} ]
+ s^2 J \int dt \sum_{I=0}^\infty  \frac{1}{2}
(\hat{{\bf n}}_{I,1} + \hat{{\bf n}}_{I,2})^2
+ \kappa \hat{{\bf n}}_{I,2} \cdot \hat{{\bf n}}_{I+1,1}
+ s \int dt \sum_{I=0}^\infty {\bf B} \cdot (\hat{{\bf n}}_{I,1}
- \hat{{\bf n}}_{I,2}) ~,
\label{dsc-action}
\end{equation}
where $\Omega[{\bf \hat n}]$ is the solid-angle subtended by
${\bf \hat n}$. This can be written as,
\begin{equation}
\Omega[\hat{{\bf n}}] = \int dt \, \int_{0}^{1} du \,
\hat{{\bf n}} \cdot \partial_t {\bf n} \times
\partial_u\hat{{\bf n}}.
\end{equation}
Here $u$ is a fictitious dimension such that ${\bf \hat n}(u,t)|_{u=1}
= {\bf \hat n}(t)$ and ${\bf \hat n}(u,t)|_{u=0}={\bf \hat z}$.

Similarly we obtain in the dual representation,
\begin{equation}
\label{S-dual}
\tilde{S}  = \tilde{S}_{edge} + \tilde{S}_{bulk},
\end{equation}
where,
\begin{eqnarray}
\tilde{S}_{edge} &=&is \Omega [ \hat{{\bf n}}_{E} ]
+ s^2  \tilde{J} \int dt  \tilde{\kappa} \hat{\bf n}_E \cdot
\hat{\bf n}_{0,1}
+ s \int dt {\bf B_0} \cdot \hat{{\bf n}}_{E}, \\
\tilde{S}_{bulk} &=& is \sum_{I=0}^\infty  \Omega
[ \hat{{\bf n}}_{I,1} ] +
\Omega [ \hat{{\bf n}}_{I,2} ]
+ s^2 \tilde{J} \int dt \sum_{I=0}^\infty \frac{1}{2}
(\hat{{\bf n}}_{I,1} + \hat{{\bf n}}_{I,2})^2
+ \tilde{\kappa} \hat{{\bf n}}_{I,2} \cdot \hat{{\bf n}}_{I+1,1}
\nonumber \\
&&- s \int dt \sum_{I=0}^\infty {\bf B} \cdot (\hat{{\bf n}}_{I,1}
- \hat{{\bf n}}_{I,2}).
\end{eqnarray}

\subsection{Change of variables}
Suppressing, for the time being, the index $I$ then the new dimer
field variables ${\bf m}$ and ${\bf l}$ are defined as follows
\begin{eqnarray}
{{\bf m}} &=& \frac{(\hat{{\bf n}}_1 - \hat{{\bf n}}_2)}{2},\\
{\bf l} &=& \frac{(\hat{{\bf n}}_1 + {\hat{\bf n}}_2)}{2} .
\end{eqnarray}
Here the variable $\bf m$ describes the quantum fluctuations of
the {\em anti-ferromagnetic} ordering whereas $\bf l$ is
associated with a {\em ferromagnetic} ordering of the spin chain.
Since one expects the former to control the physics of the problem,
the idea next is to eliminate the $\bf l$ in a standard manner and
formulate an effective action in terms of the field variable
$\bf m$ alone. It is easy to see that the effective theory only
depends on the vector fields $\hat{\bf m}$ with unit length. To
show this, notice that,
\begin{eqnarray}
{\hat{\bf n}_1}^2 &=& m^2+l^2 + 2 {{\bf m}}.{\bf l} = 1,
\nonumber \\
{\hat{\bf n}_2}^2 &=& m^2+l^2 - 2 {{\bf m}}.{\bf l} = 1.
\end{eqnarray}
Since we also have ${{\bf m}} \cdot {\bf l}=0$, it follows that
$ m^2  =  1 - l^2$. Therefore, up to quadratic order in $l$, the
vector fields $\hat{{\bf n}}_1$, $\hat{{\bf n}}_2$ can be
written as,
\begin{eqnarray}
\label{n1}
\hat{{\bf n}}_1 & = & ~~\bigg(1-\frac{l^2}{2}\bigg) \hat{{\bf m}}
+ {\bf l},
\\
\label{n2}
\hat{{\bf n}}_2 & = & -\bigg(1-\frac{l^2}{2}\bigg) \hat{{\bf m}}
+ {\bf l}.
\end{eqnarray}
Using Eqs.~(\ref{n1}) and (\ref{n2}) we now rewrite the terms in
the action in terms of the new set of variables $\hat{{\bf m}}_I$
and ${\bf l}_I$.
We proceed by presenting the final answer for
$S_{eff}$, the intermediate steps can be found in Appendix A.
This is obtained by integrating over the Gaussian
fluctuations in the $\bf l$ fields (which amounts to a systematic
expansion of the theory in powers of $1/s$) and after taking the
continuum limit. The result is,
\begin{eqnarray}
S_{eff} = \int_0^\infty dx ~ \int_0^\beta dt ~ \mathcal{L},
\end{eqnarray}
where,
\begin{eqnarray}
\label{bulk}
\mathcal{L} &=& \frac{is\kappa}{(1+\kappa)}
\hat{{\bf m}} \cdot \partial_t\hat{{\bf m}} \times \partial_x
\hat{{\bf m}}
 + \frac{\kappa Js^2a}{2(1+\kappa)}
 \partial_x\hat{{\bf m}} \cdot \partial_x\hat{{\bf m}}
+ \frac{1}{2(1+\kappa)Ja}
 \partial_t\hat{{\bf m}} \cdot \partial_t\hat{{\bf m}}
- \frac{s}{a} {\bf B} \cdot \hat{\bf m}.
\end{eqnarray}
Here, $a/2$ is the lattice spacing. It is important to remark that
the theory is defined with {\em free boundary conditions} on the
vector field variables $\hat{\bf m}$. Notice that the presence of
the staggered magnetic field ($\bf B$) term eliminates all the
arbitrariness of the problem, indicating that $S_{eff}$ is the
appropriate quantum theory for anti-ferromagnetic ordering.
Next we compare the results with those obtained in the dual
representation. In this case we obtain an extra piece associated
with the edge vector field $\hat{\bf n}_{E}$. The answer can
be written as,
\begin{equation}
\tilde{S}_{eff} = \tilde{S}_0 +
\int_0^\infty dx \int_0^\beta dt~ \tilde{\mathcal{L}}
\end{equation}
where,
\begin{equation}
\label{edge1}
\tilde{S}_{0}  =  is~ \Omega[\hat{{\bf m}}(0,t)],
\end{equation}
\begin{equation}
\label{bulk1}
\tilde{\mathcal{L}}  =
-\frac{is\tilde{\kappa}}{(1+\tilde{\kappa})}
\hat{{\bf m}} \cdot \partial_t\hat{{\bf m}} \times
\partial_x\hat{{\bf m}}
+ \frac{\tilde{\kappa} \tilde{J} s^2a}{2(1+\tilde{\kappa})}
 \partial_x\hat{{\bf m}} \cdot \partial_x\hat{{\bf m}}
+ \frac{1}{2(1+\tilde{\kappa})\tilde{J} a}
 \partial_t\hat{{\bf m}} \cdot \partial_t\hat{{\bf m}}
- \frac{s}{a} {\bf B} \cdot \hat{\bf m},
\end{equation}
and $\hat{\bf m}(0,t) =\hat{\bf n}_E(t)$.
\subsection{Nonlinear sigma model}
By rewriting the edge piece $\tilde{S}_0$
\begin{equation}
\tilde{S}_0 = is \Omega[{\bf m}(0)] = is \int dt \int dx~
\hat{{\bf m}} \cdot \partial_t \hat{{\bf m}} \times \partial_x
\hat{{\bf m}}
\end{equation}
we conclude that $S_{eff}$ and the dual theory $\tilde{S}_{eff}$
are identically the same. Thus we obtain the Lagrangian for the
standard $O(3)$ nonlinear sigma model,
\begin{eqnarray}
\mathcal{L}_{NLSM} & = & \frac{1}{2g}\bigg[c\,
\partial_x\hat{{\bf m}} \cdot \partial_x\hat{{\bf m}} + \frac{1}{c}
\partial_t\hat{{\bf m}} \cdot \partial_t\hat{{\bf m}}\bigg]
+ i \frac{\theta}{4\pi}
\hat{{\bf m}} \cdot \partial_t \hat{{\bf m}} \times \partial_x
\hat{{\bf m}} - \frac{s}{a} {\bf B} \cdot \hat{\bf m}.
\label{hal-dsc}
\end{eqnarray}
We have introduced the spin-wave velocity $c$, the coupling constant
$g$ and the instanton angle $\theta$ which are expressed as follows,
\begin{eqnarray}
\label{hal-dsc-c}
c & = & ~as~~~\sqrt{\kappa} J
~~~~~~= ~as~~~\sqrt{\tilde{\kappa}} \tilde{J}, \\
\label{hal-dsc-g}
g & = & s^{-1}~ \frac{(1+\kappa)}{\sqrt{\kappa}}
~~~~=~ s^{-1}~ \frac{(1+\tilde{\kappa})}
{\sqrt{\tilde{\kappa}}}, \\
\label{hal-dsc-th}
\theta & = & 4\pi s ~~ \frac{\kappa}{1+\kappa}
~~~~~=~ 4\pi s \left[1 - \frac{\tilde{\kappa}}
{1+\tilde{\kappa}} \right].
\end{eqnarray}
\comment{
In the dual theory we obtain,
\begin{eqnarray}
{\tilde c} & = & as~~~\sqrt{\tilde{\kappa}} \tilde{J}, \\
{\tilde g} & = & s^{-1}~ \frac{(1+\tilde{\kappa})}
{\sqrt{\tilde{\kappa}}},\\
{\tilde \theta} & = & 4\pi s \left[1 - \frac{\tilde{\kappa}}
{1+\tilde{\kappa}} \right].
\end{eqnarray}
The two theories are related to each other by the dual
transformation (the interchange of strong and weak bonds),
\begin{eqnarray*}
\kappa &=& \frac{1}{\tilde{\kappa}}, \\
J &=& \tilde{J} \tilde{\kappa}.
\end{eqnarray*}
}
\subsection{Validity of the approximations}
\label{valid}
There are two standard approximations made in the Haldane mapping.
First that ${\bf l}$ is small and second that ${\bf m}$ is a slowly
varying field on the length scale of the lattice spacing.
The small ${\bf l}$ approximation (i.e retaining terms up to $l^2$)
is basically a {\em kinematic} one that replaces the dimer by a rotor.
For a single dimer, it results in an action
\begin{equation}
S=\int dt \frac{1}{2J}\partial_t{\bf \hat m}\cdot\partial_t{\bf \hat m}.
\end{equation}
The Hilbert space of the rotor consists of one representation of
each spin-$j$, $j=0,1,2,....\infty$ the spectrum being
$E_j=\frac{1}{2J}j(j+1)$. The spin-$s$ dimer has an identical Hilbert
space and spectrum except that it is truncated at $j=2s$. Thus the
approximation becomes exact as $s\rightarrow\infty$. However, as we
will argue, we may also expect it to be good at energy scales
$E \le J$, at small $\kappa(\tilde \kappa)$. At $\kappa(\tilde \kappa)=0$,
the lowest excitations for both the decoupled rotor and the decoupled dimer
system {\em at all s}, consist of one rotor(dimer) being excited to the
triplet state. For weak coupling, the low energy physics of both the
systems is that of weakly dispersing, weakly interacting triplets.
Therefore even at $s=\frac{1}{2}$, we may expect the small ${\bf l}$
approximation to be good for at small $\kappa(\tilde \kappa)$ for
energy scales $E\le J$.
The second approximation consist of taking the ``naive" continuum
limit of the coupled rotor system which we call the {\em rotor
chain} (RC). Namely assuming that the low energy physics is
correctly described by slowly varying ${\bf m}$ fields. This is
clearly not valid for the physics of the triplets, since the $\bf
m$ fields are weakly coupled. However, if at energy scales $E <<
J$ the triplets are decoupled from the edge spin, then they are
irrelevant to the physics at these scales. Thus if the system,
under renormalization, flows to $\kappa=0$, the naive continuum
approximation will not affect the physics at energy scales small
compared to the triplet gap.
\section{Geometries}
\begin{figure}
\psfrag{eps}[l,b][l,b]{$\epsilon$}
\psfrag{ef}[l,b][l,b]{$\epsilon_F$} \psfrag{E}[c,b][l,b]{$S_E$}
\begin{center}
\includegraphics[width=240pt]{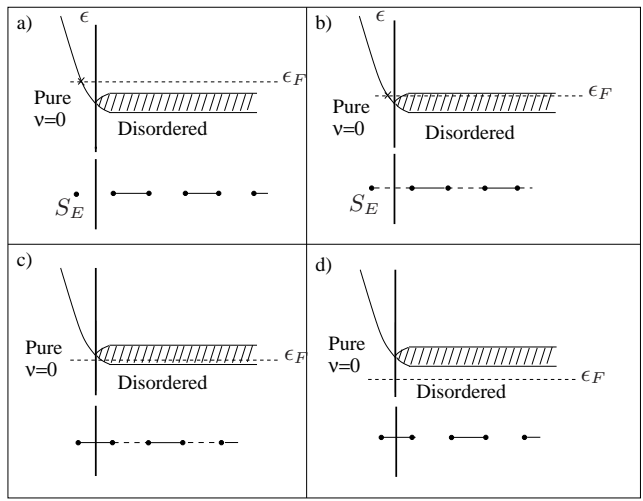}
\caption{\label{seminf} Energy spectrum for a semi-infinite
  quantum Hall systems versus $x$ and the corresponding dimerized spin chain.
  In $(a)$ we sketch the situation of a fully occupied Landau band
  $\nu=1$ (i.e. $\epsilon_F$ lies above the lowest Landau states) with massless
  excitations at the edge.
  The corresponding DSC consists of completely decoupled
  dimers ($\kappa=\infty$) with an isolated dangling spin at the edge, $S_E$.
  In $(b)$ and $(c)$ we sketch the situation where
  $\epsilon_F$ lies within the band such that $1>\nu>1/2$ (corresponding to
  the presence of weakly coupled edge spin) and $0<\nu<1/2$
  (corresponding to the absence of a dangling edge spin) respectively.
  In situation $(d)$ we have
  $\nu=0$ ($\epsilon_F$ lies below the Landau band) and the
  corresponding DSC has completely decoupled dimers only ($\kappa=0$).}
\end{center}
\end{figure}
In this Section we discuss the correspondence between quantum Hall
systems of various different geometries and dimerized
$s=\frac{1}{2}$ spin chains. This correspondence is established by
comparing the parameters entering the effective nonlinear $\sigma$
model action on the basis of which one expects that both systems
display the same (super universal) features. The imaginary time
$t$ in the spin system translates into periodic boundary
conditions the $y$ direction (or a {\em cylinder geometry}) in the
quantum Hall system. Furthermore, by considering a partially
occupied lowest Landau band with a filling fraction $0 \leq \nu
\leq 1$ then the correspondence is given by~\cite{Prange}
\begin{eqnarray}\label{sigmaxx}
\sigma_{xx} &\longleftrightarrow& 1/2g =
\frac{\sqrt{\kappa}}{1+\kappa}, \\\label{sigmaxy}
\sigma_{xy} = \nu
&\longleftrightarrow& \frac{\theta}{2\pi} =
\frac{\kappa}{1+\kappa}.
\end{eqnarray}
Here, $\sigma_{xx}$ and $\sigma_{xy}$ are the mean field values of
the dimensionless {\em longitudinal} and {\em Hall} conductance
respectively. Notice that a varying dimerization parameter
$\kappa$ in the DSC has the same meaning as a varying Fermi energy
$\epsilon_F$ or filling fraction $\nu$ in the quantum Hall system.
In what follows we shall represent the spin chain by an array of
spins along the $x$ axis and compare it with the disordered Landau
level system at zero temperature that is depicted in the plane of
energy versus the $x$ axis.
\subsection{Semi-infinite systems}

In Fig.~\ref{seminf} we sketch the semi-infinite quantum Hall
system with an edge at $x=0$ and the corresponding DSC. The ``edge
states" in the quantum Hall system are represented by the center
of the cyclotron orbits that formally lie outside the sample
($x<0$). When the Fermi level ($\epsilon_F$) lies above the lowest
landau band (Fig.~\ref{seminf}a) then the bulk of the system is
gapped but there are gapless excitations at the edge $x=0$. In the
effective non-linear $\sigma$ model action the bare parameters
$\sigma_{xx}$ and $\sigma_{xy} = \nu$ are $0$ and $1$
respectively. According to Eqs.~(\ref{hal-dsc})-(\ref{hal-dsc-th})
this corresponds to a system of completely decoupled dimers with a
dangling spin at the edge ($S_E$). Next, in Fig.~\ref{seminf}b we
consider the case where $\epsilon_F$ lies within the Landau band
such that $1 > \nu > 1/2$. We now have $\sigma_{xx}>0$ and
$1/2<\sigma_{xy}<1$. The corresponding DSC has a weak bond at the
edge (dashed line). Similarly, for $0< \nu < 1/2$ (Fig.
\ref{seminf}c) we have $\sigma_{xx}>0$ and $0<\sigma_{xy}<1/2$ and
the edge bond for the DSC is a strong one (bold line). Finally, in
Fig.~\ref{seminf}d we show the case where $\epsilon_F$ lies below
the Landau band or $\nu = 0$. In this case we have
$\sigma_{xx}=\sigma_{xy}=0$ and there are clearly no edge states.
The corresponding DSC consists of completely decoupled dimers
without a dangling spin at the edge.

\subsection{Finite systems and DSC with even number of sites}
\begin{figure}
\psfrag{eps}[l,b][l,b]{$\epsilon$}
\psfrag{ef}[l,b][l,b]{$\epsilon_F$}
\psfrag{e1}[c,b][l,b]{$S_{E_1}$}
\psfrag{e2}[c,b][l,b]{$S_{E_2}$}
\begin{center}
\includegraphics[width=240pt]{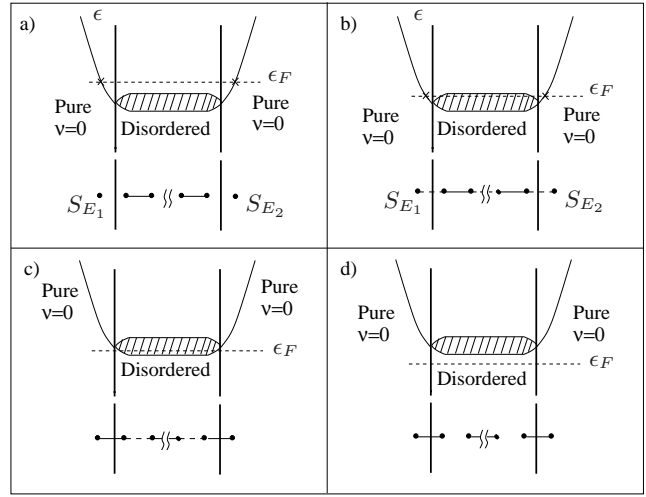}
\caption{\label{evenfinit} Energy spectrum for Finite quantum Hall
  systems and correspondence with DSC's with an even number of
  sites. In $(a)$ $\epsilon_F$ lies above the lowest Landau band and
  the DSC consisting of decoupled dimers has isolated spins at both
  the edges. In $(b)$ and $(c)$ $\epsilon_F$ is within the band with
  $\nu>1/2$ (weakly coupled edge spins at both the edges) and
  $\nu<1/2$ (no dangling edge spin) respectively.  In $(d)$
  $\epsilon_F$ lies below the Landau band and the corresponding DSC
  has completely decoupled dimers with no isolated spins at the edges.}
\end{center}
\end{figure}
A finite disordered sample has edges at $x=0$ and $x=L$
respectively (Fig.~\ref{evenfinit}). It is easy to see that the
corresponding spin chain is finite as well but it must have an
{\em even} number of sites. This is so because the quantum Hall
system with a completely filled Landau band ($\nu = 1$,
Fig.~\ref{evenfinit}a) has massless excitations at both the edges. The
corresponding DSC must therefore have dangling spins at both the
edges and this only happens when the total number of sites is {\em
even}. For $\nu = 0$ (Fig.~\ref{evenfinit}d) there are no quantum
Hall edge states and the corresponding spin chain has dimers at
both the edges (bold lines), i.e. no dangling spins. The
intermediate situations $1/2 < \nu < 1$ and $0 < \nu < 1/2$ are
depicted in Figs.~\ref{evenfinit}c and \ref{evenfinit}d respectively.
\subsection{Finite systems and DSC with odd number of sites}
\begin{figure}
\psfrag{E}[c,b][l,b]{$S_E$} \psfrag{eps}[l,b][l,b]{$\epsilon$}
\psfrag{ef}[l,b][l,b]{$\epsilon_F$}
\begin{center}
\includegraphics[width=240pt]{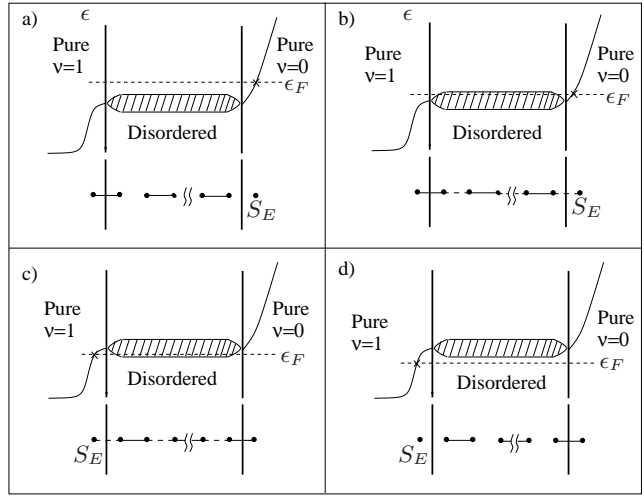}
\caption{\label{oddfinit} Energy spectrum for a {\em semi
  infinite}
  quantum Hall system with a fixed filling fractions $\nu=1$ for $x<0$
  and $\nu=0$ for $x>L$. The corresponding DSC is {\em finite} with an
  {\em odd} number of sites in the interval $0<x<L$. In situation $(a)$
  the Fermi energy $\epsilon_F$ lies above the lowest Landau band
  and the corresponding DSC consists of decoupled dimers with an dangling
  edge spin on the right. In $(b)$ and $(c)$ $\epsilon_F$ lies within the
  band with $\nu>1/2$ (corresponding to a weakly coupled edge spin on the right)
  and $\nu<1/2$ (corresponding to weakly coupled edge spin on the left)
  respectively. In $(d)$ we show the case where
  $\epsilon_F$ lies below the Landau band. The corresponding DSC
  has completely decoupled dimers with a dangling spin at the left edge.}
\end{center}
\end{figure}
Fig.~\ref{oddfinit} sketches the correspondence between a quantum
Hall system and a finite spin chain with an {\em odd} number of
sites. In this case the quantum Hall system defined for $0<x<L$ is
flanked on the right hand side ($x>L$) by the trivial vacuum or
$\nu=0$. On the left hand side ($x<0$), however, the system is
flanked by a semi-infinite quantum Hall state with a fixed filling
fraction $\nu=1$ such that in this region the Fermi level
$\epsilon_F$ is always located above the Landau band. The effect
of the $\nu=1$ quantum Hall state for $x<0$ on the energy of the
massless edge excitations at $x=0$ is indicated in
Fig.~\ref{oddfinit}. Notice that this system formally describes a
semi-infinite quantum Hall system defined for $-\infty < x < L$
where, say, the value of the external magnetic field for $x<0$ is
different from that for $x>0$. The Lagrangian for $0<x<L$ has
parameters $\sigma_{xx}$ and $\sigma_{xy}$ as in
Eqs.~\eqref{sigmaxx} and ~\eqref{sigmaxy}. However, for $x<0$ the
action is not zero but, rather, the parameters $\sigma_{xx}$ and
$\sigma_{xy}$ are now fixed and given by $0$ and $1$ respectively.
It is readily verified that the quantum Hall system with varying
values of $\epsilon_F$ or $\nu$ in the interval $0<x<L$ displays
the same basic features as the DSC with an {\em odd} number of
sites and with varying values of the parameter $\kappa$, see
Figs.~\ref{oddfinit} a-d. The Lagrangian in this case is defined by
\begin{eqnarray}
\mathcal{L}_{NLSM} & = &  \frac{i}{2} \hat{{\bf m}} \cdot
\partial_t \hat{{\bf m}} \times \partial_x \hat{{\bf m}}~~~~(x<0)
\\
& = & \frac{1}{2g}\bigg[c\,
\partial_x\hat{{\bf m}} \cdot \partial_x\hat{{\bf m}} + \frac{1}{c}
\partial_t\hat{{\bf m}} \cdot \partial_t\hat{{\bf m}}\bigg]
+ \frac{i\theta}{4\pi} \hat{{\bf m}} \cdot \partial_t
\hat{{\bf m}} \times \partial_x \hat{{\bf m}}~~(0<x<L)
\end{eqnarray}

\section{The real space renormalization group scheme}
We now formulate a real space renormalization group scheme using a
combination of Hamiltonian and path integral techniques.  The
scheme involves a systematic perturbative expansion in the weak
inter-dimer coupling and in time derivatives. We compute the RG
equations to lowest non-trivial order in both the coupling and in
derivatives and discuss the solutions.

\subsection{The decimation scheme}
We begin with the path integral representation of the DSC,
\begin{eqnarray}
Z&=&tr~ e^{-\beta H}
\label{sppint}
=\int \prod_{I\alpha}{\cal D}[{\bf {\hat n}}_{I\alpha}]
e^{-S[\{ {\bf {\hat n}}_{I\alpha} \}]},
\end{eqnarray}
where the action is given in Eq.~\eqref{bulk}. The decimation
scheme consists of integrating out half the dimers (say the odd
ones) in the path integral in Eq.~\eqref{sppint} to obtain an
effective action for the even dimers,
\begin{equation}
\label{seff1}
S^{eff}=
\sum_{I\alpha}-\frac{i}{2}\Omega[{\bf {\hat n}}_{2I\alpha}]
+\frac{1}{4}\int_{-\infty}^{\infty} d\tau \sum_I \left(
{\bf \hat n}_{I1}.{\bf \hat n}_{I2} \right)
+S^{eff}_{int},
\end{equation}
where $S^{eff}_{int}$ is given by,
\begin{eqnarray}
\nonumber e^{-S^{eff}_{int}}&=&\int \prod_{I\alpha}{\cal D}[{\bf \hat
n}_{(2I+1),\alpha}]  \exp \left(-\sum_{I\alpha}-\frac{i}{2}\Omega[{\bf \hat
n}_{(2I+1),\alpha}]\right)\\
&&~~~~~~~~~~~~~~~~~~~~\times \exp \left(-\frac{1}{4}
\int_{-\infty}^{\infty} d\tau \sum_I {\bf \hat n}_{(2I+1),1}
\cdot {\bf \hat n}_{(2I+1),2}\right)
\nonumber \\ &&~~~~~~~~~~~~~~~~~~~~\times
 \exp \left(-\frac{1}{4}\int_{-\infty}^{\infty} d\tau \sum_I
\kappa\left[{\bf \hat n}_{(2I,2)} \cdot {\bf \hat n}_{(2I+1),1}
+ {\bf \hat n}_{(2I+1),2} \cdot {\bf \hat n}_{(2I+2),1}\right] \right),
\nonumber \\
e^{-S^{eff}_{int}} &=&\prod_I\int{\cal D}[{\bf \hat n}_1]{\cal D}
[{\bf \hat n}_2]
\exp \left(\frac{i}{2}\sum_\alpha\Omega[{\bf \hat n}_\alpha]\right)
\exp \left(-\frac{1}{4}
\int_{-\infty}^{\infty}d\tau~ \left({\bf \hat n}_1.{\bf \hat n}_2
+\kappa~ \left[{\bf \hat n}_2.{\bf \hat n}_{(2I+2),1}
+{\bf \hat n}_1.{\bf \hat n}_{2I,2}\right]\right)\right).
\label{seffint}
\end{eqnarray}
Therefore, computing $S^{eff}_{int}$ involves computing the
partition function of a two-spin system in the presence of time
dependent external fields. This two-spin problem can be expressed
in the Hamiltonian formalism as,
\begin{equation}
e^{-F[{\bf m}_1,{\bf m}_2]}=tr \left( T \exp \left[-\int_{-\infty}^{\infty}
d\tau~ (h_0+\kappa h_{int}) \right] \right),
\end{equation}
where,
\begin{eqnarray}
\label{2sph0}
h_0&=&\frac{1}{2}({\bf S}_1+{\bf S}_2)^2,\\
\label{2sphint}
h_{int}&=&{\bf S}_1.{\bf m}_2(\tau)+{\bf S}_2.{\bf m_1}(\tau).
\end{eqnarray}
Here we have defined,
\begin{equation}
\label{m12def}
{\bf m}_1=\frac{1}{2}{\bf \hat n}_{(I-1),1},~~~
{\bf m}_2=\frac{1}{2}{\bf \hat n}_{(I+1),2}.
\end{equation}

\subsection{The renormalization group equations}

$F$ can be expanded systematically in a cumulant expansion in powers
of $\kappa$. To order $\sim\kappa^2$,
\begin{equation}
\label{F1}
F=-\frac{\kappa^2}{2!}\int_{-\infty}^{\infty} d\tau_1 d\tau_2
\langle 0\vert T(h_{int}(\tau_1)h_{int}(\tau_2))\vert 0\rangle.
\end{equation}
This is computed to be,
\begin{equation}
\label{F2}
F= -\frac{\kappa^2}{4}\int_{-\infty}^{\infty} d\tau
\int_{0}^{\infty}dt e^{-t}m^a(\tau+t/2)m^a(\tau-t/2),
\end{equation}
where ${\bf m}= {\bf m}_1 - {\bf m}_2$.
For fields slowly varying over a time scale of $\sim 1$, we can expand
$m^a(\tau\pm t)$ about $\tau$ and develop a local derivative expansion
for $F$. For the moment we neglect the derivatives and get,
\begin{equation}
\label{floc}
F = -\frac{\kappa^2}{8}\int_{-\infty}^{\infty}d\tau~{\bf m_1}.{\bf m_2}.
\end{equation}
Using Eqs.~(\ref{seff1},\ref{seffint},\ref{m12def},\ref{floc}) and
relabeling the sites $2I\rightarrow I$, we get the effective Hamiltonian,
\begin{equation}
\label{effham}
H_{eff}=\sum_I \left( {\bf S}_{I1}.{\bf S}_{I2}
+\kappa^\prime {\bf S}_{I2}.{\bf S}_{I+1 1} \right).
\end{equation}
This is exactly of the initial form in Eq.~(\ref{dcham}), with
the renormalized coupling $\kappa^\prime = \kappa^2/2$. We therefore have
the recursion relation for the coupling constant,
\begin{equation}
\label{recrel0}
\kappa^{n+1} = \frac{(\kappa^{n})^2}{2}.
\end{equation}
The solution to this recurrence relation is,
\begin{equation}
\label{rrsol0}
\kappa^{n} = 2\left(\frac{\kappa^0}{2}\right)^L,~~L \equiv 2^n.
\end{equation}
Thus, there are two fixed points,
\begin{equation}
\label{fp2}
\kappa^*=0~~{\rm and}~~\kappa^*=2,
\end{equation}
and the coupling constant flows from $\kappa^*=2$ to $\kappa^*=0$.
Note that the Hamiltonian in Eq.~(\ref{dcham}) with coupling
$\kappa =1$, the $s=1/2$ uniform Heisenberg chain, is known to be
gap-less. This implies that $\kappa^*=1$, corresponding to $\theta = \pi$
has to be a fixed point of the RG equations. As we will see in the next
section, this error is due to the neglecting of the time derivatives.

\subsection{Time derivatives}
\label{tder}

The RHS of Eq.~\eqref{F2} can be expanded to second order in the
derivatives and the effective action computed. We obtain,
\begin{eqnarray}
\nonumber
S^{eff}&=&\int_{-\infty}^{\infty} d\tau
\left( \sum_{I\alpha}
-\frac{i}{2}{\bf A}_{mon}({\bf \hat n}_{I\alpha})\cdot
\partial_\tau {\bf \hat n}_{I\alpha}
+\frac{\kappa^\prime}{8}\partial_\tau{\bf \hat n}_{I\alpha}
\cdot \partial_\tau {\bf \hat n}_{I\alpha}
+\sum_I \frac{1}{8}{\bf \hat n}_{I1}\cdot {\bf \hat n}_{I2}
+ \frac{\kappa^\prime}{4} \left[
{\bf \hat n}_{I2} \cdot (1+\partial_\tau^2){\bf \hat n}_{I+12} \right]
\right).
\label{seffder}
\end{eqnarray}
The system described by the above action is no longer a spin chain.
It now corresponds to a system of charged particles confined to the
surface of a sphere with a unit monopole at the center. If we neglect
the derivative terms coupling neighbouring sites, the Hilbert
space at each site has one representation of all half odd integer
angular momenta, $J=1/2, 3/2, 5/2, ...$. Namely, at every site we have
a fermionic rotor. Thus under renormalization, the spin chain goes over
to a fermionic rotor chain (FRC). This can be understood qualitatively as
follows. The degree of freedom at each site in the effective action
corresponds to the motion of the original spin at that site and a block of
even number of spins that have been integrated out. Thus we can expect many
excitations with higher (half odd integer) angular momenta.

\subsection{The fermionic rotor chain}

We therefore begin with a general FRC described by the action,
\begin{eqnarray}
S = \int_{-\infty}^{\infty} d\tau\, \Bigg[& \sum_{I\alpha} & \left(
-\frac{i}{2}{\bf A}_{mon}({\hat {\bm \xi}}_{I\alpha})\cdot
\partial_\tau {\hat {\bm \xi}}_{I\alpha}
+\frac{\kappa_3 {s^\prime}^2}{2} ~ \partial_\tau {\hat {\bm \xi}}_{I\alpha}\cdot
\partial_\tau {\hat {\bm \xi}}_{I\alpha} \right) \nonumber\\
+ &\sum_I&  {s^\prime}^2 \left(\kappa_4 {\hat {\bm \xi}}_{I1} \cdot
\partial_\tau^2{\hat {\bm \xi}}_{I2}
+\kappa_2 {\hat {\bm \xi}}_{I2} \cdot
\partial_\tau^2{\hat {\bm \xi}}_{(I+1),1} \right)
+\sum_I{s^\prime}^2 \left( {\hat {\bm \xi}}_{I1} \cdot {\hat {\bm \xi}}_{I2}
+\kappa_1 {\hat {\bm \xi}}_{I2} \cdot {\hat {\bm \xi}}_{(I+1),1} \right) \Bigg].
\label{genac}
\end{eqnarray}
We have introduced four coupling constants, $\kappa_1$, $\kappa_2$,
$\kappa_3$ and $\kappa_4$. $s^\prime\equiv 3/2$ for reasons that will
become clear later. A model of this type for a uniform chain has been
previously considered in reference \cite{rands}. Our dimerized model has
the duality of the DSC. If we make the transformation,
\begin{equation}
{\hat {\bm \xi}}_{I1}\rightarrow{\hat {\bm \xi}}_{I2},~~~
{\hat {\bm \xi}}_{I2}\rightarrow{\hat {\bm \xi}}_{(I+1),1}~~~
\mbox{and}~\tau\rightarrow\frac{\tau}{\kappa_1},
\label{dutrans}
\end{equation}
we get back exactly the same model with $\kappa_i\rightarrow{\tilde \kappa_i}$,
where,
\begin{equation}
{\tilde \kappa_1}=\frac{1}{\kappa_1},~~~
{\tilde \kappa_2}=\kappa_4\kappa_1,~~~
{\tilde \kappa_3}=\kappa_3\kappa_1~~~\mbox{and}~
{\tilde \kappa_4}=\kappa_2\kappa_1.
\label{frcdual}
\end{equation}
The action for the dimer problem is now,
\begin{equation}
S = \int_{-\infty}^{\infty} d\tau\, \left[ \sum_{\alpha} \left(
-\frac{i}{2} {\bf A}_{mon}({\hat {\bm \xi}}_{\alpha}) \cdot
\, \partial_\tau {\hat {\bm \xi}}_{\alpha}
+\frac{\kappa_3 {s^\prime}^2}{2}~ \partial_\tau {\hat {\bm \xi}}_{\alpha}
\cdot \partial_\tau {\hat {\bm \xi}}_{\alpha}
+ {s^\prime}^2~{\hat {\bm \xi}}_{\alpha}.\,{\bf m}_{\alpha} \right)
+\kappa_4 {s^\prime}^2 {\hat {\bm \xi}}_{I1} \cdot
\partial_\tau^2{\hat {\bm \xi}}_{I2}
+{s^\prime}^2  {\hat {\bm \xi}}_{1} \cdot  {\hat {\bm \xi}}_{2}\right],
\label{dimac}
\end{equation}
where,
\begin{eqnarray}
\label{m12}
{\bf m}_1&=&(\kappa_1+\kappa_2~\partial_\tau^2)~
{\hat {\bm \xi}}_{(I\!-\!1),2},\\
{\bf m}_2&=&(\kappa_1 + \kappa_2~\partial_\tau^2)~
{\hat {\bm \xi}}_{(I\!+\!1),1}.
\end{eqnarray}
We first put $\kappa_4=0$ and discuss the effects of it being
non-zero later. The Hamiltonian corresponding to the action in
Eq.~\eqref{dimac} is then,
\begin{eqnarray}
\label{dimham}
h_d &=& h_0 + h_{int},\\
\label{dimham0}
h_0 &=& \frac{1}{2\kappa_3{s^\prime}^2} \sum_\alpha
({{\bf J}_\alpha} \cdot {{\bf J}_\alpha} - \frac{1}{4}),\\
\label{dimhamint}
h_{int} &=& {s^\prime}^2 {\hat {\bm \xi}}_1.{\hat {\bm \xi}}_2 + s^\prime \sum_\alpha
{\hat {\bm \xi}}_\alpha.{\bf m}_\alpha.
\end{eqnarray}
${\bf J}_\alpha$ are the angular momentum operators.
The spectrum of $h_0$ is,
\begin{eqnarray}
\label{dh0spect}
h_0\vert(j_1,m_1),(j_2,m_2)\rangle&=&E_{j_1,j_2}
\vert(j_1,m_1),(j_2,m_2)\rangle,\\
\nonumber
E_{j_1,j_2}&=& \frac{1}{2\kappa_3{s^\prime}^2}
\sum_\alpha \left(j_\alpha(j_\alpha+1)-\frac{1}{4}\right),
\end{eqnarray}
with $j_\alpha=1/2,3/2...$ and $m_\alpha = -j_\alpha ... j_\alpha$
The ground states are the four $j_1=j_2= 1/2$ states.
When $\kappa_3 <<1$, the gap between the $(1/2,1/2)$ states and the
excited states is very large. In this limit, to a very good approximation,
we can project the model into
the $(j_1,j_2)=(1/2,1/2)$ subspace. The matrix elements of ${\bf\hat n}$
between
the $j=1/2$ states can be explicitly computed using the monopole harmonics
and we have,
\begin{equation}
\langle {\scriptstyle \frac{1}{2}},\sigma_1 \vert n^a \vert
{\scriptstyle \frac{1}{2}},\sigma_2 \rangle
={\frac{2}{3}} S^a = {\frac{1}{s^\prime}} S^a .
\end{equation}
Thus to leading order in $\kappa_3$, we recover the two-spin
system as the effective Hamiltonian,
\begin{equation}
\label{heff}
  h_{eff} = {\bf S}_1.{\bf S}_2 + \sum_\alpha {\bf S}_\alpha \cdot
{\bf m}_\alpha . \nonumber
\end{equation}
The previous results can now be used to integrate out the dimer system and
get the  recursion relations to second order in the $\kappa$'s. They are,
\begin{eqnarray}
\label{rec1}
  \kappa_1^{\left(n+1\right)} &=& \frac{ {\left(
\kappa_1^{\left(n\right)}\right)}^2}{2} ,\\
\label{rec2}
  \kappa_2^{\left(n+1\right)} &=& \frac{ {\left(
\kappa_1^{\left(n\right)}\right)}^2}{2} +
 \kappa_1^{n}  \kappa_2^{n} ,\\
\label{rec3}
\kappa_3^{\left(n+1\right)} &=&   \kappa_3^{n} +
\frac{ {\left(\kappa_1^{\left(n\right)}\right)}^2}{2} +
 \kappa_1^{n}  \kappa_2^{n} .
\end{eqnarray}

\subsection{The strong coupling fixed points}

The recursion relations in Eqs.~\eqref{rec1}, \eqref{rec2} and
\eqref{rec3} can be solved explicitly with the initial conditions,
\begin{equation}
\label{incond}
\kappa_1^{(0)}=\kappa_{10},~~\kappa_2^{(0)} = \kappa_{20},
~~\kappa_3^{(0)} = \kappa_{30} .
\end{equation}
The solution is,
\begin{eqnarray}
\label{sol1}
  \kappa_1^{\left(n\right)} &=& 2\left(\frac{\kappa_{10}}{2}
\right)^{L} , \\
\label{sol2}
 \kappa_2^{\left(n\right)} &=& \left(aL-1\right)
\kappa_1^{(n)} , \\
\label{sol3} \kappa_3^{\left(n\right)} &=&
\kappa_{30}+\sum_{m=0}^{n-1} \kappa_2^{\left(n\right)} .
\end{eqnarray}
Where $L \equiv 2^n$ as before and
$a=(1+\kappa_{10}/\kappa_{20})$. These recursion relations
have an infrared stable coupling fixed point which can be written
as,
\begin{equation}
\label{FP} \kappa_1=0,~~\kappa_2=0,~~\kappa_3=\kappa_3^*
\neq\kappa_{30} .
\end{equation}
As will be discussed in more detail below, the above fixed point
is obtained provided the initial condition $\kappa_{10} < 2$ is
satisfied. Notice that when $\kappa_{10}=2$ we have $\kappa_1^{n}
=2$ but the value of $\kappa_2^{n}$ diverges for large $n$ and so
does $\kappa_3^(n)$ as well as $\kappa_3^*$. Thus while the
intermediate coupling fixed point, $\kappa_1^*=2$ obtained earlier
remains a fixed point of Eq.~\eqref{rec1}, the value of the new
parameters lie outside the range of validity of the recursion
relations, $\kappa_2^*=\kappa_3^*=\infty$. In particular, the
decimation procedure has been evaluated under the assumption
$\kappa_3 <<1$ which is clearly violated by the intermediate
coupling fixed point with $\kappa_1^*=2$. In what follows we shall
study the validity of our renormalization procedure in more
detail. More specifically, we are interested in finding a subset
of the parameter space $(\kappa_1 , \kappa_2 , \kappa_1 )$ that
satisfies $\kappa_3^{(n)} <<1$ for all positive integer values of
$n$.

\begin{figure}
\includegraphics[height=250pt]{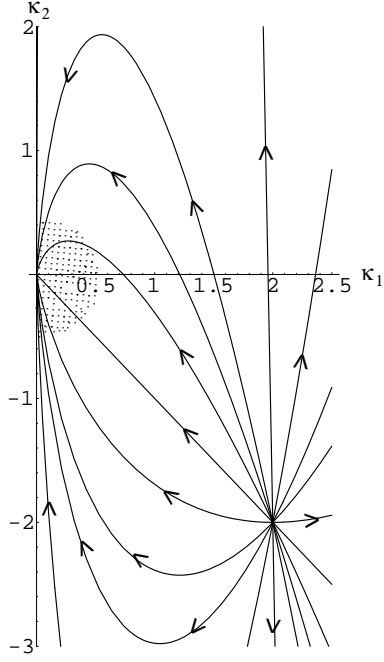}
\caption{\label{rgflow} Renormalization group flow in the
$\kappa_1$- $\kappa_2$ plane.}
\end{figure}

To start we write the recursion relations in differential form.
>From Eq.~\eqref{sol1} and \eqref{sol2} we immediately obtain the
following renormalization group equations,
\begin{eqnarray}
\label{beta1}
\beta_1 (\kappa_1 ) &=& \frac{d\kappa_1}{d\ln L} = \kappa_1
\ln \kappa_1 /2 , \\
\label{beta2} \beta_2 (\kappa_1 ,\kappa_2 ) &=&
\frac{d\kappa_2}{d\ln L} = \kappa_1 + \kappa_2 +\kappa_2 \ln
\kappa_1 /2 .
\end{eqnarray}
The various different renormalization group trajectories in the
($\kappa_1 , \kappa_2$) plane can be found be solving the
differential equation,
\begin{equation}
\label{DV}
\frac{d\kappa_2}{d\kappa_1} = \beta_2 (\kappa_1
,\kappa_2 ) / \beta_1 (\kappa_1 ) .
\end{equation}
The various solutions to Eqs.~\eqref{beta1} - \eqref{DV} are
sketched in Fig.~\ref{rgflow}. This clearly indicates that the
fixed point $(\kappa_1 , \kappa_2) = (0,0)$ is a completely {\em
stable} in the {\em infrared}. On the other hand, a {\em finite}
intermediate fixed point does exist at $(\kappa_1 , \kappa_2) =
(2,-2)$ but it is completely {\em unstable}.

The main issue next is to obtain an explicit expression for the
quantity $\kappa_3^*$ in Eq.~\eqref{FP}. Let $( \kappa_1^{(n)} ,
\kappa_2^{(n)})$ denote a point close to the strong coupling fixed
point $(0,0)$. It is then possible to find a function $F$,
\begin{equation}
\kappa_3^{(n)} = F(\kappa_1^{(n)} , \kappa_2^{(n)}) ,
\end{equation}
that is the solution of the recursion relation of Eq.~(\ref{rec3}).
To lowest order in the quantities $\kappa_1^{(n)} ,
\kappa_2^{(n)}$ we obtain,
\begin{equation}
F(\kappa_1^{(n)} , \kappa_2^{(n)}) = \kappa_3^*
-\frac{(\kappa_1^{(n)})^2}{2} - \kappa_1^{(n)} \kappa_2^{(n)}
+{\cal O} \left( (\kappa^{(n)})^4 \right).
\end{equation}
On the basis of this result we obtain the following expressions
for $\kappa_3^{(n)}$ with $n=0$ and $n \rightarrow \infty$
respectively,
\begin{eqnarray}
\kappa_3^{(0)} & = & \kappa_3^*  -\frac{(\kappa_1^{(0)})^2}{2} -
\kappa_1^{(0)} \kappa_2^{(0)} +{\cal O} \left( (\kappa^{(0)})^4
\right) , \\
\label{largen} \kappa_3^{(n)} & \rightarrow & \kappa_3^* + {\cal
O} \left( e^{-2L/\xi} \right).
\end{eqnarray}
The first of these equations expresses $\kappa_3^*$ in terms of
the renormalization points $\kappa^{(0)}$.
\begin{equation}\label{kappastar}
\kappa_3^* = \kappa_3^{(0)} + \frac{(\kappa_1^{(0)})^2}{2} +
\kappa_1^{(0)} \kappa_2^{(0)} +{\cal O} \left( (\kappa^{(0)})^4
\right).
\end{equation}
Eq.~ \eqref{largen}, on the other hand, tells us that the quantity
$\kappa_3^*$ appearing in Eqs.~(\ref{kappastar}) and (\ref{FP}) are
indeed identically the same. The final result of
Eq.~\eqref{kappastar} can be used to identify the set of points
$\left\{ \kappa_1^{(0)} , \kappa_2^{(0)} \right\}$ in the
$\kappa_1 , \kappa_2$ plane for which the conditions,
\begin{equation}
0 \leq \kappa_3^{(0)} \ll 1 , ~~~0 \leq \kappa_3^* \ll 1 ,
\end{equation}
can be satisfied simultaneously. This set is indicated by the
shaded area in Fig.~\ref{rgflow}.

We will now discuss the effects of turning on $\kappa_4$. The
first point to note is that $\kappa_4$ does not renormalize. This
is because in the action in Eq.~(\ref{genac}) the $I^{th}$
dimer couples to ${\hat {\bm \xi}}_{I+12}~{\rm and}~{\hat {\bm \xi}}_{I-1,1}$.
Thus, in the absence of next nearest neighbor couplings,
integrating out the $I^{th}$ dimer cannot produce a coupling
between the two ${\hat {\bm \xi}}$'s on the same dimer.

The $\kappa_4$ term breaks the $SU(2)\times SU(2)$ of the dimer
Hamiltonian $h_0$ in Eq.~\eqref{dimham0}. For small $\kappa_4$,
the low energy subspace will remain the $(j_1,j_2)=(1/2,1/2)$
subspace. Then the two-spin effective Hamiltonian has to be of the
form,
\begin{equation}
\label{heffpr}
  h_{eff} = \left(1+\gamma(\kappa_4)\right){\bf S}_1.{\bf S}_2
  + \sum_\alpha {\bf S}_\alpha.  \,{\bf m}_\alpha,
\end{equation}
where $\gamma(0)=0$. A non-zero $\gamma$ does not change the RG
flow discussed above qualitatively. Its main effect is just to
change the value of the intermediate coupling fixed point.
However, as we discussed earlier, our current leading order
calculation is not expected to be accurate in that regime. Thus
$\kappa_4$ does not effect the strong coupling fixed point
significantly and we will not discuss its effect in any more
detail at this stage.

\section{Haldane mapping for the fermionic rotor chain}
\label{nlsmfrc}
We will now examine the FRC in some detail. The low energy physics
of the FRC can be mapped on to the NLSM similar to the Haldane
mapping of the DSC to the NLSM. Consider the FRC defined in
Eq.~\eqref{genac}. As we did previously, define,
\begin{equation}
\label{mvec}
{\bf m} = \frac{{\hat{\bm \xi}}_1 - {\hat {\bm \xi}}_2}{2}
~~~\mbox{and} ~~
{\bm \lambda} = \frac{{\hat{\bm \xi}}_1 + {\hat {\bm \xi}}_2}{2} .
\end{equation}
They satisfy the following constraints,
\begin{equation}
m^2+\lambda^2 = 1 ~~~\mbox{and}~~
{\bf m}\cdot{\bm \lambda} = 0 .
\end{equation}
We now express the different terms in the action in terms of the
dimer variables ${\bf m}$ and ${\bm \lambda}$ defined in
Eqs.~\eqref{mvec}. Then,
\begin{equation}
-\frac{i}{2}\sum_\alpha{\bf A}_{mon}({\hat {\bm \xi}}_{I\alpha})
= -i {\bm \lambda_I}\cdot{{\bf \hat m}}_I\times\partial_\tau{{\bf \hat m}}_I.
\end{equation}
This is an exact expression \cite{naveen}. It is not possible to write
the sum of the two solid angle terms as a local $\tau$ integral
because the coordinate ${{\bf \hat m}}$ is singular when $\vec\lambda=0$.
However these points correspond to the two fermionic rotors in the
dimer being aligned parallel and are hence not important configurations.
The rest of the terms are expanded to quadratic order
in $|{\bm \lambda}_I|$. Then,
\begin{eqnarray}
S&=&\int_{-\infty}^{\infty}d\tau~ \Big(
{s^\prime}^2 (\kappa_2+\kappa_3+\kappa_4)
\partial_\tau{{\bf \hat m}}_I\cdot\partial_\tau{{\bf \hat m}}_I
-\kappa_1{s^\prime}^2{{\bf \hat m}}_I\cdot{{\bf \hat m}}_{(I+1)}\Big)
+~S_{\lambda},
\label{vrcac} \\
S_\lambda&=&\int_{-\infty}^{\infty}d\tau~
\left({\bm \lambda}_I\cdot{\cal A}_{IJ}{\bm \lambda}_J+{\bm \lambda}_I
\cdot{\vec {\cal B}}_I \right) \\
\nonumber
{\cal A}_{IJ}&\equiv& \left(2{s^\prime}^2 - {s^\prime}^2
(\kappa_3- \kappa_2 - \kappa_4) \partial_\tau^2 \right)
\delta_{IJ} + \kappa_1{s^\prime}^2~\delta_{(I+1)J} \nonumber \\
&& + \left( \kappa_1{s^\prime}^2{{\bf \hat m}}_I\cdot{{\bf \hat m}}_{(I+1)}
+ {s^\prime}^2 (\kappa_2+\kappa_3+\kappa_4)
~\partial_\tau{{\bf \hat m}}_I\cdot\partial_\tau{{\bf \hat m}}_I
\right) \delta_{IJ}
\nonumber \\
&& \\
\nonumber
{\vec{\cal B}}_I&\equiv&
\kappa_1{s^\prime}^2({{\bf \hat m}}_{(I+1)}-{{\bf \hat m}}_{I-1})
-i {{\bf \hat m}}_I\times\partial_\tau{{\bf \hat m}}_I
\end{eqnarray}

\subsection{The continuum limit}

We now take the continuum limit as follows. The continuum fields,
${\hat {\bf m}}(\tau,x)$ and {\boldmath $\lambda$}$(\tau,x)$ are defined as,
\begin{eqnarray}
{\bf {{\bf \hat m}}}(\tau,aI)&\equiv&{{\bf \hat m}}_I(\tau) , \\
{\bm \lambda}(\tau,aI)&\equiv&{\bm \lambda}_I(\tau) ,
\end{eqnarray}
where $a$ is the lattice spacing. We now assume that the fields
${{ \bf \hat m}}(\tau,x)$ and ${\bm \lambda}(\tau,x)$ vary slowly
over a length scale of $a$ and a time scale of 1, and hence drop all
derivatives of order greater than two. Then we have,
\begin{eqnarray}
\nonumber
S=\int_{-\infty}^{\infty}d\tau dx~&&\left(
\frac{{\cal I}}{2}~ \partial_\tau{{\bf \hat m}}
\cdot\partial_\tau{{\bf \hat m}}
+\frac{a\kappa_1{s^\prime}^2}{2}~
\partial_x{{\bf \hat m}}\cdot\partial_x{{\bf \hat m}}\right)
+S_{\lambda},
\\\nonumber
S_\lambda = \int_{-\infty}^{\infty}d\tau dx~ & \Bigg[ &
\frac{\mu}{2}~ \partial_\tau{\bm \lambda}\cdot
\partial_\tau{\bm \lambda}-\frac{\kappa_1{s^\prime}^2 a}{2}~
\partial_x{\bm \lambda}\cdot\partial_x{\bm \lambda}
\nonumber \\
+&&~\left(\frac{2{s^\prime}^2(1+\kappa_1)}{a}
+ \frac{{\cal I}}{2}
\partial_\tau{{\bf \hat m}}\cdot\partial_\tau{{\bf \hat m}}
-\frac{\kappa_1 {s^\prime}^2 a}{2}
\partial_x{{\bf \hat m}}\cdot\partial_x{{\bf \hat m}}\right)
{\bm \lambda}\cdot{\bm \lambda} \nonumber \\
+&&~(2\kappa_1 {s^\prime}^2 ~ \partial_x{{\bf \hat m}}
-\frac{i}{a} {{\bf \hat m}} \times \partial_\tau{{\bf \hat m}} )
\cdot {\bm \lambda} \Bigg],
\end{eqnarray}
where ${\cal I} \equiv (2{s^\prime}^2/a) ( \kappa_2 +
\kappa_3 + \kappa_4)$ and $\mu \equiv (2{s^\prime}^2/a) (
\kappa_3 - \kappa_2 - \kappa_4)$. We now do the rescaling
$\lambda\rightarrow \sqrt{a} \lambda$, and drop terms of ${\cal
O}(a)$. Then we have,
\begin{equation}
S_\lambda = \int_{-\infty}^{\infty}d\tau dx~\left[
2{s^\prime}^2(1+\kappa_1)~ {\bm \lambda}\cdot{\bm \lambda}
+\left(2\kappa_1 {s^\prime}^2 ~ \partial_x{{\bf \hat m}}
-\frac{i}{a} {{\bf \hat m}} \times \partial_\tau{{\bf \hat m}}\right)
\cdot \sqrt{a} {\bm \lambda} \right].
\end{equation}
Integrating out ${\bm \lambda}$ we obtain the effective NLSM.
\begin{equation}
S_{NLSM} = \int d\tau dx~\left( \frac{{\tilde {\cal I}}}{2}
~\partial_\tau{{\bf \hat m}}\cdot\partial_\tau{{\bf \hat m}}
+\frac{\gamma}{2}~ \partial_x{{\bf \hat m}}\cdot
\partial_x{{\bf \hat m}} + \frac{i \theta}{4\pi}~
{{\bf \hat m}}\cdot\partial_x{{\bf \hat  m}}\times
\partial_\tau{{\bf \hat m}} \right) , \label{NLSM}
\end{equation}
where the moment of inertia ${\tilde {\cal I}}$, the stiffness constant
$\gamma$ and the instanton angle $\theta$ are given by,
\begin{eqnarray}
\label{mominerfrc}
 {\tilde {\cal I}}&=&\left( {\cal I}
 +\frac{1}{4{s^\prime}^2 a (1+\kappa_1)}\right) , \\
 \gamma&=&\frac{a\kappa_1 {s^\prime}^2}{(1+\kappa_1)} , \\
 \theta&=&2\pi\frac{\kappa_1}{(1+\kappa_1)} .
\label{thetaprime}
\end{eqnarray}
The coupling constant $g$ and spin wave velocity $c$ are then,
\begin{eqnarray}
\label{1bygfrc}
\frac{1}{g}&=& \left( \frac{\kappa_1(8{s^\prime}^4(1+\kappa_1)
 (\kappa_2+\kappa_3 +\kappa_4) +1)}{4(1+\kappa_1)^2}
 \right)^{\frac{1}{2}} , \\
 c&=& \left(\frac{4 a^2 \kappa_1
 {s^\prime}^4} {8{s^\prime}^4(1+\kappa_1)(\kappa_2+\kappa_3 +\kappa_4)
 +1} \right)^{\frac{1}{2}} .
\end{eqnarray}

\subsection{The weak coupling regime}

In the Haldane mapping of the DSC to the NLSM, the weak coupling
regime of the NLSM corresponds to large $s$ spin chains. Thus for
the spin $\frac{1}{2}$ system, we cannot access this regime. In the
FRC, however, this regime corresponds to the
$\kappa_3\sim\kappa_2\sim\kappa_4>>1$ region. As we can see from
equations (\ref{mominerfrc}) and (\ref{1bygfrc}), the moment of
inertia becomes very large and $g$ very small.

When the moment of inertia is very large, we expect the system to
behave semi-classically. To make this explicit, we can scale the
euclidean time in Eq.~(\ref{genac}), $\tau\rightarrow \sqrt{
\kappa_3}\tau$. The action is then written as,
\begin{eqnarray}
\nonumber
S= -\int_{-\infty}^{\infty} d\tau\,\Bigg[&&\left(\sum_{I\alpha}
 \frac{i}{2}{\bf A}_{mon}({\hat {\bm \xi}}_{I\alpha}) \cdot
 \partial_\tau {\hat {\bm \xi}}_{I\alpha}
 +\frac{\sqrt{\kappa_3}{s^\prime}^2}{2}~
 \partial_\tau {\hat {\bm \xi}}_{I\alpha} \cdot\partial_\tau
{\hat {\bm \xi}}_{I\alpha} \right)
 \\
 &&+\sum_I \left( \frac{\kappa_4}{\kappa_3}~
 {\hat {\bm \xi}}_{I1}\cdot\partial_\tau^2{\hat {\bm \xi}}_{I2}
 +\frac{\kappa_2}{\kappa_3}~
 {\hat {\bm \xi}}_{I2} \cdot \partial_\tau^2
{\hat {\bm \xi}}_{(I+1),1}
+~{\hat {\bm \xi}}_{I1}\cdot {\hat {\bm \xi}}_{I2}
+~\kappa_1~ {\hat {\bm \xi}}_{I2}\cdot{\hat {\bm \xi}}_{(I+1),1} \right)
\Bigg]
\label{genacsemi}
\end{eqnarray}
The derivation of the NLSM presented
in section (\ref{nlsmfrc}) can thus be looked upon as the leading
order results of a systematic semi-classical expansion,
$1/\sqrt{\kappa_3}$ being the expansion parameter.

\subsection{Quantization of the Hall conductance}
The above mapping of the fermionic rotor chain onto the nonlinear
$\sigma$ model holds at each step of the renormalization group
transformation. Hence, following Eqs.~\eqref{thetaprime} and
\eqref{1bygfrc} we can associate a different set of NLSM
parameters $g^{(n)} , \theta^{(n)}$ with each set of coupling
constants $\kappa^{(n)}_i$ that is defined by the decimation
procedure. However, for each step $(n)$ in the decimation
procedure the vector field variable $\hat m$ is defined for a
different lattice constant ${a^{(n)}} = 2^n (2a) = L(2a)$ or
momentum scale $\lambda^{(n)} = \frac{\pi}{a^{(n)}}$. To discuss
the limit of scaling we must consider $L \rightarrow \infty$ or,
equivalently, large values of $n$. Specifying to the case
$s^\prime =1/2$ then the quantities $1/ g^{(n)}$ and
$\theta^{(n)}$ for large values of $n$ approach the fixed point
values $1/g =0 $ and $\theta =2\pi m$ arbitrary closely and are
related to one another according to,
\begin{equation}\label{asymflow}
 \left( \frac{1}{g^{(n)}} \right)^2 =
 \left( \kappa_3^* + 1 \right) {\left| \frac{\theta^{(n)}}{2\pi} - m
\right|} \propto  e^{-2L/\xi} .
\end{equation}
Here, $m=0,1$ and $\kappa_{3}^*$ is given by
Eq.~\eqref{kappastar}. Notice that the asymptotic form of
Eq.~\eqref{asymflow} defines a different parabola in the $1/g$,
$\theta$ coupling constant plane for each different value of
$\kappa_3^*$.

Next, in order to make contact with the quantum Hall effect it is
convenient to re-express the results of Eq.~\eqref{NLSM} in terms
of the Grassmannian field variable $Q$.
\begin{equation}
S_{NLSM} = \frac{1}{8} \int_{0}^{\infty} dx
 \int^{\infty}_{-\infty} d\tau \left[ \sigma_{xx}^{(n)}~
 \mbox{tr}~\left( \partial_\mu Q \partial_\mu Q\right) +
\sigma_{xy}^{(n)}~\mbox{tr}~\left(\epsilon_{\mu\nu} Q \partial_\mu Q
 \partial_\nu Q\right) \right],
\end{equation}
where,
\begin{eqnarray} \label{asymflow1}
 \sigma_{xx}^{(n)} &=&  \frac{1}{g^{(n)}}  = {\mathcal O}(e^{-L/\xi}), \\
 \sigma_{xy}^{(n)} &=&  \frac{\theta^{(n)}}{2\pi} = m +{\mathcal O} (
 e^{-2L/\xi}) .
\end{eqnarray}
Thus at energy scales much less than 1 (note that we are working
with $J=1$) the low energy effective action is precisely equal to
the one dimensional action for massless chiral edge excitations,
Eq.~\eqref{edscac}, with $k$ now denoting the {\em quantized Hall
conductance} $\sigma_{xy}^{(n)} =  \frac{\theta^{(n)}}{2\pi} = m$.
For the semi-infinite system as considered here this one
dimensional action is defined along the edge located at $x=0$.

\section{Summary and conclusion}

We have introduced and investigated the concept of super
universality in quantum Hall and quantum spin liquids. This
concept has arisen from earlier studies of the $U(N+M)/U(N)\times
U(M)$ NLSM's in the physical context of the quantum Hall effect.

The observability and the extreme flatness of the quantum Hall
plateaus correspond to the existence of infrared stable fixed
points at $\theta=2\pi k$ as well as the existence of gapless
chiral edge excitations at these points. The transitions between
adjacent quantum Hall plateaus are generally described by an
unstable fixed point located at $\theta=\pi (2k+1)$ which
corresponds to a diverging correlation length in the system.

Recent work on the large $N$ $CP^{N-1}$ models have explicitly
demonstrated the above super universal features of the instanton
vacuum~\cite{PruiskenBaranovVoropaev}. In particular, the subtle
issue of the existence of a diverging correlation length at
$\theta=\pi$ despite the transition being first order was
clarified.

In this work we have focused on the physics of the strong
coupling fixed points at $\theta=2\pi k$. We first examined the
theory in the ``bare" strong coupling limit and demonstrated the
existence of gapless chiral edge excitations. The edge
correlations are independent of the ultraviolet cutoff procedure
and also independent of $M$ and $N$.

We then further concentrated on the $M=N=1$ case corresponding to
the $O(3)$ NLSM. We showed that the edge correlations map on to
the correlations of a free spin at the edge. This motivated us to
re-examine the Haldane map of DSC to the NLSM treating the edges
carefully. We showed how to associate the different geometries of
the quantum spin liquid with the quantum Hall liquid with
different boundary conditions.

Next we developed a real space renormalization group scheme for
the DSC using a combination of path integral and Hamiltonian
methods. The technique exploits the fact that under the Haldane
mapping the strong coupling regime of the NLSM corresponds to a
system of weakly coupled dimers. It is then possible to implement
the renormalization group perturbatively in the dimer-dimer
coupling. We showed that under renormalization the theory flows
away from the DSC to a more complicated theory with four
parameters, the FRC. The renormalization procedure was then
implemented for the FRC. The recursion relations were obtained to
leading order. These were solved exactly demonstrating the
existence of the strong coupling fixed points at $\theta = 0~{\rm
and}~2\pi$. The Haldane map from the FRC to the NLSM then shows
that the fixed point action of the instanton vacuum is precisely
given by the action for massless chiral edge excitations.

In summary we can say that the FRC, like the large $N$ expansion
of the $CP^{N-1}$ model, can be used as an explicit example for
demonstrating the {\em robust quantization} of the Hall
conductance. The most important results of this paper are the
scaling results of Eqs.~\eqref{asymflow} and ~\eqref{asymflow1}
since they describe the strong coupling regime in the
$\sigma_{xx}$, $\sigma_{xy}$ conductance plane that previously has
not been accessible. These results seem to have a much broader
range of validity than considered in this paper. For example,
preliminary investigations have shown that the decimation
procedure can be extended to include the general case of an
$SU(N)$ quantum spin liquid and the Grassmannian $U(2M)/U(M)
\times U(M)$ NLSM, yielding basically the same results. These
investigations as well as systematic expansion procedures in
powers of $1/s$ will be reported in separate papers ~\cite{SUN}.

\begin{figure}
\includegraphics[width = .45 \textwidth]{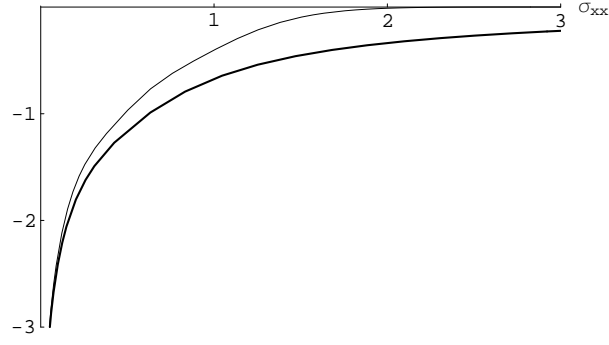}
\caption{\label{betafn} $2 \tilde{\beta}_{\sigma}$(bold curve) and
$\tilde{\beta}_{\theta}$ (dashed curve), obtained by interpolating
between the strong and weak coupling results, plotted as functions
of $\sigma_{xx}$.}
\end{figure}

\begin{figure}
\rotatebox{90}{
\includegraphics{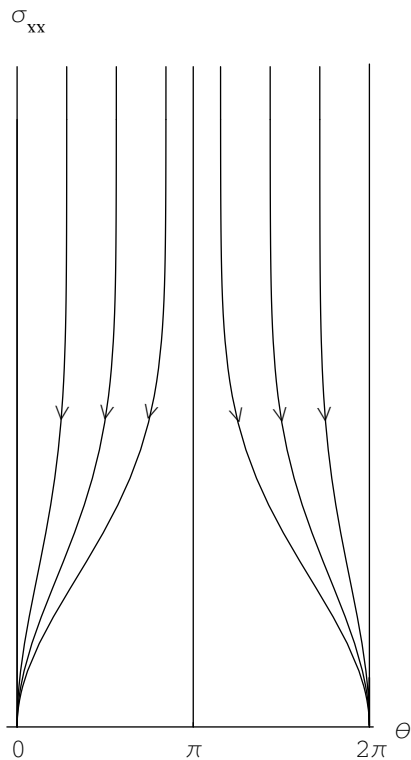}}
\caption{\label{sigtheta} Renormalization Group flow obtained from
the $\beta$ functions interpolated between the strong and weak
coupling results}
\end{figure}

To conclude this work we next present the consequences of our
results in terms of the global phase diagram for the FRC or,
equivalently, the $O(3)$ NLSM. For this purpose we shall make use
of the renormalization group results for the weak coupling regime
$g<<1$ or $\sigma_{xx} >> 1 $ that has previously been obtained
from the instanton calculations
~\cite{Pruisken,PruiskenBurmistrov}. We choose to work with the
parameters $\sigma_{xx} = 1/g$ and $\theta$. Our scaling results
of Eqs.~\eqref{asymflow} and ~\eqref{asymflow1} describing the
strong coupling regime can then be obtained as the solutions of
the following differential equations,
\begin{equation}
\label{beta-strong}
 \left. \begin{array}{l}
\tilde{\beta}_\theta \equiv \frac{\mbox{$d~\ln | \theta -2\pi m |
$}} {\mbox{$d~\ln L$}} =2
\ln \sigma_{xx} \\  \\
\tilde{\beta}_\sigma \equiv \frac{\mbox{$d~\ln \sigma_{xx}$}}
{\mbox{$d~\ln L$}} = {\mbox{$\ln \sigma_{xx}$}}
\end{array} \right\} ~\mbox{for}~ \sigma_{xx} \ll 1 .
\end{equation}
On the other hand, from the weak coupling instanton calculations
of the $O(3)$ model we have for $\theta \approx 2\pi m$
\begin{equation}
\label{beta-weak}
 \left. \begin{array}{l}
\tilde{\beta}_\theta = -2\pi D_{11} (2 \pi \sigma_{xx})^4
e^{-2\pi \sigma_{xx}} \\  \\
\tilde{\beta}_\sigma = -\left(\frac{\mbox{$1$}} {\mbox{$\pi
\sigma_{xx}$}}+\frac{\mbox{$1$}} {\mbox{$(\pi \sigma_{xx})^2$}}
\right) -2\pi D_{11} (2 \pi \sigma_{xx})^4 e^{-2\pi
\sigma_{xx}}
 \end{array} \right\}~\mbox{for}~
\sigma_{xx} \gg 1 .
\end{equation}
Here $D_{11}$ is a numerical constant which equals $0.0294$. We next
can perform a simple numerical interpolation between the asymptotic
expressions of Eqs.~(\ref{beta-strong}) and (\ref{beta-weak}) to
obtain the $\tilde{\beta}$ functions along the entire $\sigma_{xx}$
axis. This leads to the curves as shown in Fig.~\ref{betafn}. Notice
that our result for $\tilde{\beta}_\sigma$, which is otherwise well
known from the quantum theory of metals, is just the standard way of
expressing the phenomenon of {\em Anderson localization} in two
spatial dimensions.~\cite{Gangof4} The $\tilde{\beta}_\theta$ curve,
however, is a generic non-perturbative feature of the instanton vacuum
concept and fundamentally describes the formation of the quantum Hall
plateaus. From the interpolated $\tilde{\beta}$ functions shown in
Fig.~\ref{betafn} we can compute the various different renormalization
group flow lines in the $\sigma_{xx}$, $\theta$ coupling constant
plane. These different flow lines which define the {\em domain of
attraction} of the infrared ({\em quantum Hall}) fixed points
$\sigma_{xx} =0$ and $\theta=2\pi m$, are shown in
Fig.~\ref{sigtheta}. Notice that the results of this paper, notably
Fig.~\ref{sigtheta}, provide an important justification of the
originally proposed scaling diagram for the $O(3)$ NLSM as shown in
Fig.~\ref{RGflowMN}(b). While the unstable fixed point at $\theta=\pi$
in Fig.~\ref{RGflowMN}(b) cannot be accessed at the level of
approximation that we have implemented in the decimation procedure
introduced in this paper, its existence is well known and
established\cite{Zamolodchikov}. We have thus shown that quantum spin
liquids, like the quantum Hall liquid, display all the {\em super
universal} strong coupling features of an {\em instanton vacuum}.

\section{Acknowledgments}
This research was funded in part by the Dutch National Science
Foundations FOM ({\em Fundamenteel Onderzoek der Materie}) and
NWO ({\em Nederlandse organisatie voor Wetenschappelijk Onderzoek}).

\appendix

\section{}

Using the change of variables,
\begin{eqnarray}
\label{newvar1}
\hat{{\bf n}}_{I1} & = & ~~\bigg(1-\frac{l_I^2}{2}\bigg)
\hat{{\bf m}}_I + {\bf l}_I , \\
\label{newvar2}
\hat{{\bf n}}_{I2} & = & -\bigg(1-\frac{l_I^2}{2}\bigg)
\hat{{\bf m}}_I + {\bf l}_I ,
\end{eqnarray}
and expanding to quadratic order in ${\bf l}$, the action in
Eq.~\eqref{dsc-action} can be written as,
\begin{eqnarray}
S &=& \int dt~\sum_{I=1}^{\infty} \Bigg( 2s^2(1+\kappa) l_I^2
+ \kappa s^2~ \left (\hat{\bf m}_{(I+1)} - \hat{\bf m}_{I-1}
\right) \cdot {\bf l}_I
+ 2is~ \hat{\bf m}_I \times \partial_t \hat{\bf m}_I \cdot {\bf l}_I
- \kappa J s^2~ \hat{\bf m}_I \cdot \hat{\bf m}_{(I+1)} \Bigg)
\end{eqnarray}
In writing the above expression we have made the approximation
${\bf l}_I \cdot {\bf l}_{(I+1)} \approx (l_I^2 + l_{(I+1)}^2)\big/2$.
This amounts to dropping terms with more than two derivatives in the
eventual effective action for the $\hat{\bf m}$ field.
In the next step we integrate out the ${\bf l}$ field to obtain
the effective action as,
\begin{eqnarray}
S_{eff} &=& \int dt~ \Bigg[ \kappa s^2  \sum_{I=1}^{\infty}
\hat{\bf m}_I \cdot \hat{\bf m}_{(I+1)}
- \sum_{I=1}^{\infty}
\Bigg( 4i \kappa s (\hat{\bf m}_{(I+1)} - \hat{\bf m}_{I-1}) \cdot
\hat{\bf m}_I \times \partial_t \hat{\bf m}_I \nonumber \\
&&~~~~+ \kappa^2 s^2 \Big((\hat{\bf m}_{(I+1)} - \hat{\bf m}_{I-1})^2
- \left( \hat{\bf m}_{I} \cdot \left( \hat{\bf m}_{(I+1)}
- \hat{\bf m}_{I-1} \right) \right)^2 \Big)
- 4  \partial_t \hat{\bf m}_I \cdot \partial_t \hat{\bf m}_I \Bigg)
\Bigg(8 (1+ \kappa) \Bigg)^{-1} \Bigg].
\end{eqnarray}
Taking the continuum limit we get,
\begin{eqnarray}
S_{eff} &=& \int dt~dx~\Bigg[ \frac{1}{2(1+\kappa)Ja}
~ \partial_t\hat{{\bf m}}(x,t) \cdot \partial_t\hat{{\bf m}}(x,t)
+ \frac{\kappa Js^2a}{2(1+\kappa)}~
\partial_x\hat{{\bf m}}(x,t) \cdot \partial_x\hat{{\bf m}}(x,t)
\nonumber \\
&&~~~~~~~~~~~~~ +\frac{is\kappa}{(1+\kappa)}~
\hat{{\bf m}}(x,t) \cdot \partial_t\hat{{\bf m}}(x,t) \times \partial_x
\hat{{\bf m}}(x,t)   \Bigg].
\label{effec-action}
\end{eqnarray}
where $a/ 2$ is the lattice spacing.

\subsection*{Dual case}
Next we consider the dual case where there is an unpaired spin
$\tilde{{\bf n}}_E$ at the edge. As before, we first change the
variables to ${\bf m}$ and ${\bf l}$ as given by
Eqs. \eqref{newvar1} and \eqref{newvar2}.
\comment{
\begin{eqnarray}
\tilde{{\bf n}}_{I1} & = & ~~\bigg(1-\frac{l_I^2}{2}\bigg)
\tilde{{\bf m}_I} + {\bf l}_I , \\
\tilde{{\bf n}}_{I2} & = & -\bigg(1-\frac{l_I^2}{2}\bigg)
\tilde{{\bf m}}_I + {\bf l}_I .
\end{eqnarray}
}
The action in Eq.~\eqref{S-dual} can then be written as,
\begin{eqnarray}
\tilde{S} &=& \int dt~ \Bigg( is \Omega[\tilde{\bf n}_E] + \kappa J s^2
~\tilde{\bf n}_E \cdot \tilde{\bf m}_1 + \left[ \kappa s^2
~(\tilde{\bf m}_2 + \tilde{\bf n}_E) +2is~ \tilde{\bf m}_1 \cdot
\partial_t \tilde{\bf m}_1 \right] \cdot {\bf l}_1
+ s^2~ \left( 2+ \kappa - \frac{\kappa}{2} \tilde{\bf n}_E \cdot
\tilde{\bf m}_1\right) l_1^2  \nonumber \\
&&~~~~~- \kappa J s^2~ \sum_{I=1}^{\infty} \tilde{\bf m}_I
\cdot \tilde{\bf m}_{(I+1)} + \sum_{I=2}^{\infty}  \left[ \kappa s^2~
\left (\tilde{\bf m}_{(I+1)} - \tilde{\bf m}_{I-1}\right) \cdot {\bf l}_I
+ 2is~ \tilde{\bf m}_I \times \partial_t \tilde{\bf m}_I \cdot {\bf l}_I
+ 2s^2(1+\kappa)~ l_I^2 \right] \Bigg)
\end{eqnarray}
Integrating out ${\bf l}$ we obtain the effective action as,
\begin{equation}
\tilde S_{eff} = \tilde S_{edge} + \tilde S_{bulk},
\end{equation}
where,
\begin{eqnarray}
S_{edge} &=& is \Omega[\tilde{\bf n}_E] - \kappa s^2 \int dt~\Bigg[
\tilde{\bf n}_E \cdot \tilde{\bf m}_1 - \Bigg(4i \kappa s~(\tilde{\bf m}_2
+ \tilde{\bf n}_E) \cdot \tilde{\bf m}_1 \times \partial_t \tilde{\bf m}_1
\nonumber \\
&&~~~+ \kappa^2 s^2 \left[(\tilde{\bf m}_2 + \tilde{\bf n}_E)^2
- \left(\tilde{\bf m}_1 \cdot \left( \tilde{\bf m}_2 + \tilde{\bf n}_E \right)
\right)^2 \right] - 4  \partial_t \tilde{\bf m}_1 \cdot
\partial_t \tilde{\bf m}_1 \Bigg) \left(8+ 4\kappa - 2\kappa~
\tilde{\bf n}_E \cdot \tilde{\bf m}_1 \right)^{-1} \Bigg], \\
S_{bulk} &=& \int dt~ \Bigg[ \kappa s^2  \sum_{I=1}^{\infty}
\tilde{\bf m}_I \cdot \tilde{\bf m}_{(I+1)}
- \sum_{I=2}^{\infty}
\Bigg( 4i \kappa s (\tilde{\bf m}_{(I+1)} - \tilde{\bf m}_{I-1}) \cdot
\tilde{\bf m}_I \times \partial_t \tilde{\bf m}_I \nonumber \\
&&~~~~+ \kappa^2 s^2 \Big((\tilde{\bf m}_{(I+1)} - \tilde{\bf m}_{I-1})^2
- \left( \tilde{\bf m}_{I} \cdot \left( \tilde{\bf m}_{(I+1)}
- \tilde{\bf m}_{I-1} \right) \right)^2 \Big)
- 4  \partial_t \tilde{\bf m}_I \cdot \partial_t \tilde{\bf m}_I \Bigg)
\Bigg(8 (1+ \kappa) \Bigg)^{-1} \Bigg].
\end{eqnarray}
The real part of $S_{eff}$ is minimized by the choice
$\tilde{{\bf m}}_I(t)=-\tilde{\bf n}_E$, where we are free to choose
$\tilde{\bf n}_E$. Therefore small fluctuations about the minimum can be
described by the slowly varying continuous field $\tilde{\bf m}(x,t)$ with
the boundary condition $\tilde{\bf m}(0,t) = -\tilde{\bf n}_E(t)$.
\comment{
Then
\begin{equation}
is \Omega[\tilde{\bf m}(0,t)] = is~\int dt~dx~\tilde{\bf m} \cdot
\partial_t \tilde{\bf m} \times \partial_x \tilde{\bf m} .
\end{equation}
}
In the continuum limit, $\tilde{\bf m}(x,t) = -{\bf m}(x,t)$.
Then we obtain the effective dual-action as,
\begin{eqnarray}
\tilde S_{eff} = \tilde S_{edge} + is~ \Omega[\hat{\bf m}(0)]
+ \int dt~dx~&\Bigg[& \frac{1}{2(1+\kappa)Ja}~\partial_t
\hat{{\bf m}}(x,t) \cdot \partial_t\hat{{\bf m}}(x,t)
+ \frac{\kappa Js^2a}{2(1+\kappa)}~
\partial_x\hat{{\bf m}}(x,t) \cdot \partial_x\hat{{\bf m}}(x,t)
\nonumber \\ && - is~\frac{\kappa}{(1+\kappa)}~
\hat{{\bf m}}(x,t) \cdot \partial_t\hat{{\bf m}}(x,t) \times \partial_x
\hat{{\bf m}}(x,t)  \Bigg].
\label{dualeffec-action}
\end{eqnarray}
where,
\begin{eqnarray}
\tilde S_{edge} = (4+3\kappa)^{-1} \int dt~
&\bigg[& 2~ \partial_t\hat{{\bf m}}(0,t) \cdot \partial_t\hat{{\bf m}}(0,t)
-2(\kappa sa)^2~ \partial_x\hat{{\bf m}}(0,t) \cdot
\partial_x\hat{{\bf m}}(0,t) \nonumber \\
&&+4is \kappa a~
\hat{{\bf m}}(0,t) \cdot \partial_t\hat{{\bf m}}(0,t) \times \partial_x
\hat{{\bf m}}(0,t) \bigg].
\end{eqnarray}
$S_{edge}$ is a line integral along the boundary of the space-time and
therefore can be written as a bulk integral by applying Stoke's
theorem. But this will only result in higher derivative terms and hence
can be dropped altogether.

\end{document}